\documentclass[twocolumn,aps,superscriptaddress,longbibliography,amsmath,amssymb,reprint]{revtex4-1} 
\usepackage{bm}
\usepackage{amsmath}
\usepackage{amssymb}
\usepackage{times}
\usepackage[usenames,dvipsnames]{color}
\usepackage[table]{xcolor}
\usepackage{graphicx}
\usepackage{dcolumn}
\usepackage{simplewick}
\usepackage{hyperref}

\usepackage{dcolumn}
\usepackage[utf8]{inputenc}
\usepackage[T1]{fontenc}
\usepackage{mathptmx}
\DeclareUnicodeCharacter{2010}{-}

\hypersetup{
  colorlinks=false,
  colorlinks=true,
  citecolor=blue,
  linkcolor=blue,
  urlcolor=blue}

\begin{document}
\title{RCC calculation of electric dipole polarizability and correlation energy 
of Cn, Nh$^+$ and Og: Correlation effects from lighter to 
superheavy elements}

\author{Ravi Kumar}
\affiliation{Department of Physics, Indian Institute of Technology, 
             Hauz Khas, New Delhi 110016, India}

\author{S. Chattopadhyay}
\affiliation{Department of Physics, Kansas State University,  
             Manhattan, Kansas 66506, USA
             }

\author{D. Angom}
\affiliation{Department of Physics, Manipur University, 
             Canchipur 795003, Manipur, India}
\affiliation{Physical Research Laboratory,
             Ahmedabad - 380009, Gujarat,
             India}

\author{B. K. Mani}
\email{bkmani@physics.iitd.ac.in}
\affiliation{Department of Physics, Indian Institute of Technology, 
             Hauz Khas, New Delhi 110016, India}

\begin{abstract}

We employ a fully relativistic coupled-cluster theory to calculate the 
ground-state electric dipole polarizability and electron correlation energy
of superheavy elements Cn, Nh$^+$ and Og. To assess the trend of electron 
correlation as function of $Z$, we also calculate the correlation energies for 
three lighter homologs--Zn, Cd and Hg; Ga$^+$, In$^+$ and Tl$^+$; Kr, Xe and 
Rn--for each superheavy elements. The relativistic effects and quantum 
electrodynamical corrections are included using the Dirac-Coulomb-Breit 
Hamiltonian with the corrections from the Uehling potential and the self-energy. 
The effects of triple excitations are considered perturbatively in the theory. 
Furthermore, large bases are used to test the convergence of results.
Our recommended values of polarizability are in good agreement with previous 
theoretical results for all SHEs. From our calculations we find that 
the dominant contribution to polarizability is from the valence electrons
in all superheavy elements. Except for Cn and Og, we observe a decreasing 
contribution from lighter to superheavy elements from the Breit interaction. 
For the corrections from the vacuum polarization and self-energy, we observe a 
trend of increasing contributions with $Z$. From energy calculations, we 
find that the second-order many-body perturbation theory overestimates the 
electron correlation energy for all the elements considered in this work.
\end{abstract}

\maketitle
\section{Introduction}

The study of  superheavy elements (SHEs) is a multidisciplinary research area 
which provides a roadmap to investigate and understand several properties 
related to physics and chemistry \cite{turler-13,matthias-15,pershina-15,
peter-15,eliav-15,giuliani-19}. There is however a lack of experimental data 
on atomic properties of SHEs due to various challenges, such as low 
production rate, short half-lives of elements and the lack of the state of 
the art one-atom-at-a-time experimental facility, associated with atomic 
experiments \cite {turler-13,schadel-14,hoffman-06}. Moreover, the properties 
of SHEs can not be predicted based on lighter homologs, as they often differ 
due to relativistic effects in SHEs {\bf \cite{eliav-15}}. In such cases, the 
theoretical investigations of physical and chemical properties provide an 
important insight to the properties of SHEs. Moreover, the benchmark data on 
these properties from accurate theoretical predications is important for 
future experiments. Calculating accurate properties of SHEs is, however, a 
difficult task. The reason for this could be attributed to the competing 
nature of the relativistic and correlation effects in these systems. For a 
reliable prediction of the properties of SHEs both of these effects should be 
incorporated at the highest level of accuracy. In addition, large basis 
sets should be used to obtain the converged properties results.

The electric dipole  polarizability, $\alpha$, of an atom or ion is a key 
parameter which  used to probe several fundamental as well as technologically 
relevant properties in atoms and ions \cite{khriplovich-91,griffith-09,
udem-02,lewenstein-94,anderson-95,karshenboim-10}. The $\alpha$ for SHEs Cn 
and Og has been calculated in previous works, 
Refs. \cite{seth-97,nash-05,dzuba-16,pershina-08a} and 
\cite{nash-05,dzuba-16, pershina-08b,jerabek-18}, respectively. Though most 
of these results are using the CCSD(T), there is a large variation in the 
reported values for both Cn and Og. For example, the value of $\alpha$ 
reported in CCSD(T) calculation \cite{jerabek-18} is $\approx$ 25\% larger 
than the similar calculation \cite{pershina-08b}. The reason for this can, 
perhaps, be attributed to the complex nature of the electron correlation and 
relativistic effects in these systems. The other point to be mentioned here 
is that, the basis used in these calculations are not large. Moreover, 
the inclusion of the contributions from the Breit interaction and QED 
corrections is crucial to obtain the accurate and reliable values of 
$\alpha$ for SHEs.

In this work, we employ a fully relativistic coupled cluster (RCC) theory based 
method to calculate the electric dipole polarizability and the electron
correlation energy of superheavy elements Cn, Nh$^+$ and Og. 
RCC is one of the most powerful many-body theories for atomic structure 
calculations as it accounts for the electron correlation to all-orders of 
residual Coulomb interaction. We have used this to calculate the many-electron 
wavefunction and the electron 
correlation energy. The effect of external electric field, in the case of $\alpha$, 
is accounted using the perturbed relativistic coupled-cluster (PRCC) 
theory \cite{chattopadhyay-12b,chattopadhyay-13b,chattopadhyay-14,
chattopadhyay-15,ravi-20}. 
One of the key merits of PRCC in the properties calculation is that it 
does not employ the sum-over-state \cite{safronova-99,derevianko-99} approach to 
incorporate the effects of a perturbation. The summation over all the possible 
intermediate states is subsumed in the perturbed cluster operators. The leading 
order relativistic effects are accounted using the four component 
Dirac-Coulomb-Breit no-virtual-pair Hamiltonian \cite{sucher-80}. And, the effects 
of Breit, QED and triples excitations in coupled-cluster are computed using 
the implementations in our previous works \cite{chattopadhyay-12b, 
chattopadhyay-13b, chattopadhyay-14, chattopadhyay-15}.

To assess the trend of various electron correlation effects from lighter
to SHEs, we also calculate the correlation energy and the contributions from
the Breit and QED corrections to $\alpha$ for three lighter homologs in each 
SHEs: Zn, Cd and Hg in group-12; Ga$^+$, In$^+$ and Tl$^+$ in group-13; and 
Kr, Xe, and Rn in group-18. In this work, however, we do not report the values 
of $\alpha$ for these lighter homologs as these have been already reported in 
our previous works \cite{chattopadhyay-15,ravi-20,chattopadhyay-12b}. Here, 
our main focus is to get deeper insight of various correlation effects as a 
function $Z$ in each of these SHEs. 
More precisely, we aimed to: accurately calculate the value of $\alpha$ and 
correlation energy of SHEs Cn, Nh$^+$ and Og using RCC and test the convergence 
of results with very large basis; 
study the electron correlation in $\alpha$ for SHEs and assess the trend from 
lighter to SHEs; and 
examine in detail the contributions from the Breit and QED corrections to 
$\alpha$ for SHEs elements and get a deeper insight to the trend of 
contributions from lighter homologs to SHEs.

The remaining part of the paper is organized into five sections. 
In Sec. II we provide a brief description of the method used in the
polarizability calculation. In Sec. III we provide the calculational details 
such as the single-electron basis and computational challenges associated 
with polarizability calculation of SHEs. In Sec. IV we analyze and discuss 
the results from our calculations. The theoretical uncertainty in our calculation 
is discussed in Sec. V. Unless stated otherwise, all the results and equations 
presented in this paper are in atomic units ( $\hbar=m_e=e=1/4\pi\epsilon_0=1$)


\section{Method of Calculation}

The ground state wavefunction of an N-electron atom or ion in relativistic
coupled-cluster (RCC) theory is
\begin{equation}
 |\Psi_0\rangle = e^{ T^{(0)}}|\Phi_0\rangle,
 \label{psi0}
\end{equation}
where $|\Phi_0\rangle$ is the Dirac-Fock reference wavefunction and $T^{(0)}$ 
is the closed-shell coupled-cluster (CC) excitation operator. It is an 
eigenfunction of the Dirac-Coulomb-Breit no-virtual-pair Hamiltonian
\begin{eqnarray}
   H^{\rm DCB} & = & \sum_{i=1}^N \left [c\bm{\alpha}_i \cdot
        \mathbf{p}_i + (\beta_i -1)c^2 - V_{N}(r_i) \right ]
                       \nonumber \\
    & & + \sum_{i<j}\left [ \frac{1}{r_{ij}}  + g^{\rm B}(r_{ij}) \right ],
  \label{ham_dcb}
\end{eqnarray}
where $\bm{\alpha}$ and $\beta$ are the Dirac matrices. The negative-energy 
continuum states of the Hamiltonian are projected out by using the 
kinetically balanced finite GTO basis sets \cite{mohanty-91,stanton-84}, 
and selecting only the positive energy states from the finite size basis 
set \cite{grant-10,grant-06}. The last two terms, $1/r_{ij} $ and 
$g^{\rm B}(r_{ij})$, are the Coulomb and Breit interactions, respectively.

The operators $T^{(0)}$ in Eq. (\ref{psi0}) are the solutions of the coupled 
nonlinear equations
\begin{widetext}
\begin{subequations}
\begin{eqnarray}
  \langle\Phi^p_a| H_{\rm N} +  \left [ H_N, T^{(0)} \right ]
   + \frac{1}{2!} \left [\left [ H_N, T^{(0)}\right], T^{(0)}\right]
   + \frac{1}{3!} \left [\left [ \left [ H_N, T^{(0)}\right ], T^{(0)}\right ],
      T^{(0)} \right ] |\Phi_0\rangle &=& 0, \;\;\;\;\;\;\;\;\; \\
  \langle\Phi^{pq}_{ab}|  H_{\rm N}
   + \frac{1}{2!} \left [\left [ H_N, T^{(0)}\right], T^{(0)}\right]
   + \frac{1}{3!} \left [\left [ \left [ H_N, T^{(0)}\right ], T^{(0)}\right ],
      T^{(0)} \right]
   + \frac{1}{4!} \left [\left [\left [\left [ H_N, T^{(0)}\right],
     T^{(0)}\right], T^{(0)}\right], T^{(0)} \right] |\Phi_0\rangle &=& 0.
\end{eqnarray}
  \label{cceq_t0}
\end{subequations}
\end{widetext}
Here, the states $|\Phi^p_a\rangle$ and $|\Phi^{pq}_{ab}\rangle$ are the
singly- and doubly-excited determinants obtained by replacing {\em one} and 
{\em two} electrons from the core orbitals in $|\Phi_0\rangle$ with virtual 
orbitals, respectively. And,
$H_{\rm N} = H^{\rm DCB} - \langle \Phi_0|H^{\rm DCB}|\Phi_0\rangle$ is
the normal-ordered Hamiltonian.

In the presence of an external electric field, $\mathbf{E}_{\rm ext}$, the 
ground state wavefunction $|\Psi_0\rangle$ is modified due to interaction 
between induced electric dipole moment $\mathbf{D}$ of the atom and 
$\mathbf{E}_{\rm ext}$. We call the modified wavefunction as the perturbed 
wavefunction, which in PRCC is defined as
\begin{equation}
 |\widetilde{\Psi}_0\rangle = e^{T^{(0)}}\left [ 1 + \lambda 
  \mathbf{T}^{(1)}\cdot \mathbf{E}_{\rm ext} \right ] |\Phi_0\rangle,
 \label{psi0-ptrb}
\end{equation}
where $\mathbf{T}^{(1)}$ is the perturbed CC operator, and $\lambda$ is a
perturbation parameter. The wavefunction $|\widetilde{\Psi}_0\rangle$ is an
eigenstate of the modified Hamiltonian 
$H_{\rm Tot} =  H^{\rm DCB} - \lambda \mathbf{D}\cdot\mathbf{E}_{\rm ext}$.
The perturbed CC operators $\mathbf{T}^{(1)}$ in Eq. (\ref{psi0-ptrb}) are 
the solutions of the linearized PRCC 
equations \cite{chattopadhyay-12a,chattopadhyay-12b,chattopadhyay-13a, 
chattopadhyay-13b, chattopadhyay-15, ravi-20}
\begin{subequations}
\begin{eqnarray}
   \langle\Phi^p_a| H_{\rm N} 
     + \left [ H_N, {\mathbf T}^{(1)}\right] |\Phi_0\rangle
   &=& \langle\Phi^p_a|\left [ {\mathbf D}, T^{(0)}\right] |\Phi_0\rangle, \\
  \langle\Phi^{pq}_{ab}|  H_{\rm N} 
   +  \left [ H_N, {\mathbf T^{(1)}} \right ] |\Phi_0\rangle
   &=& \langle\Phi^{pq}_{ab}| \left [ {\mathbf D}, T^{(0)} \right ] 
        |\Phi_0\rangle.
\end{eqnarray}
  \label{cceq_t1}
\end{subequations}
The single and double excitations in the couple-cluster theory capture most
of the correlation effects and hence, the operators  $T^{(0)}$ and 
${\mathbf T}^{(1)}$ are approximated as $T^{(0)} = T_{1}^{(0)} + T_{2}^{(0)}$
 and ${\mathbf T}^{(1)} = {\mathbf T}_{1}^{(1)} + {\mathbf T}_{2}^{(1)}$,
respectively. This is referred to as the coupled-cluster single and 
double (CCSD) approximation \cite{purvis-82}. In the present work we,
however, also incorporate the triple excitations 
perturbatively \cite{chattopadhyay-15}.

The perturbed wavefunction from Eq. (\ref{psi0-ptrb}) is used
to calculate the ground state polarizability. In PRCC 
\begin{equation}
  \alpha = -\frac{\langle \widetilde \Psi_0|{\mathbf D}|\widetilde 
         \Psi_0\rangle} {\langle \widetilde \Psi_0|\widetilde \Psi_0\rangle}.
\end{equation}
Using the expression of  $|\widetilde \Psi_0\rangle$ from Eq. (\ref{psi0-ptrb})
we can write
\begin{equation}
  \alpha = -\frac{\langle \Phi_0|\mathbf{T}^{(1)\dagger}\bar{\mathbf{D}} +
   \bar{\mathbf{D}}\mathbf{T}^{(1)}|\Phi_0\rangle}{\langle\Psi_0|\Psi_0\rangle},
   \label{dbar}
\end{equation}
where $\bar{\mathbf{D}} = e^{{T}^{(0)\dagger}}\mathbf{D} e^{T^{(0)}}$, and
in the denominator $\langle \Psi_0| \Psi_0\rangle$ is the normalization factor.
Considering the computational complexity, we truncate $\bar{\mathbf{D}}$ and
the normalization factor to second order in the cluster operators $T^{(0)}$ . 
From our previous study \cite{mani-10}, using an iteration scheme we found that
the terms with third and higher orders contribute very less to the properties.


\section{Calculational Detail}

\subsection{Single-electron basis}

In the RCC and PRCC calculations, it is crucial to use an orbital basis set 
which provides a good description of the single-electron wave functions and 
energies. In the present work, we use Gaussian-type orbitals (GTOs) 
\cite{mohanty-91} as the basis. We optimize the orbitals as well as the 
self-consistent-field energies of GTOs to match the 
GRASP2K \cite{jonsson-13} results. In the Table \ref{basis}, we provide
the values of the exponents $\alpha_0$ and $\beta$ \cite{mohanty-91}
of the occupied orbital symmetries for Cn, Nh$^+$ and Og. For further
improvement, we incorporate the effects of Breit interaction, vacuum 
polarization and the self-energy corrections. This is crucial to obtain the 
value of the dipole polarizability of SHEs accurately, where the 
relativistic effects are larger due to higher $Z$. The effects of finite 
charge distribution of the nucleus are incorporated 
using a two-parameter finite size Fermi density distribution.

In the Appendix, we compare the orbital energies of Cn (Table \ref{ene-cn}), 
Nh$^+$ (Table \ref{ene-nh}) and Og (\ref{ene-og}) with 
GRASP2K \cite{jonsson-13} and B-spline \cite{zatsarinny-16} data. 
As seen from the tables, the GTOs orbital energies are in excellent agreement 
with the numerical values from GRASP2K. The largest differences are 
$3.4\times10^{-4}$, $4.8\times10^{-4}$ and $9.2\times10^{-4}$ hartree in the 
case of $4f_{5/2}$, $1s_{1/2}$ and $2p_{1/2}$ orbitals of Cn, Nh$^+$ 
and Og, respectively. The corrections from the vacuum polarization, 
$\Delta\epsilon_{\rm Ue}$, and the self-energy, $\Delta\epsilon_{\rm SE}$, to 
the orbital energies are provided in the Table \ref{delta-en}. Our results
match well with the previous calculation \cite{karol-18b} for Cn and Og.

\begin{table*}
   \caption{The $\alpha_0$ and $\beta$ parameters of the even tempered GTO basis 
    used in our calculations.}
   \label{basis}
   \begin{ruledtabular}
   \begin{tabular}{ccccccccc}
     Atom & \multicolumn{2}{c}{$s$} & \multicolumn{2}{c}{$p$} & 
     \multicolumn{2}{c}{$d$} & \multicolumn{2}{c}{$f$} \\
     \cline{2-3} \cline{4-5} \cline{6-7} \cline{8-9}
     & $\alpha_{0}$  & $\beta$ & $\alpha_{0}$ & $\beta$  
     & $\alpha_{0}$  & $\beta$  & $\alpha_{0}$  & $\beta$ \\
    \hline
     ${\rm Cn}$  &\, 0.00545 &\, 1.870 &\,0.00475 &\, 1.952 &\, 0.00105 &\, 1.970 &\, 0.00380 &\,1.965 \\
     ${\rm Og}$  &\, 0.00410 &\, 1.910 &\,0.00396 &\, 1.963 &\, 0.00305 &\, 1.925 &\, 0.00271 &\,1.830 \\
     ${\rm Nh^+}$ &\, 0.05200 &\, 1.912 &\,0.03650 &\, 1.655 &\, 0.05550&\, 1.945 &\, 0.00455 &\,1.945 \\ 
   \end{tabular}
   \end{ruledtabular}
\end{table*}

\begin{figure*}
  \includegraphics[height=17cm, angle=-90]{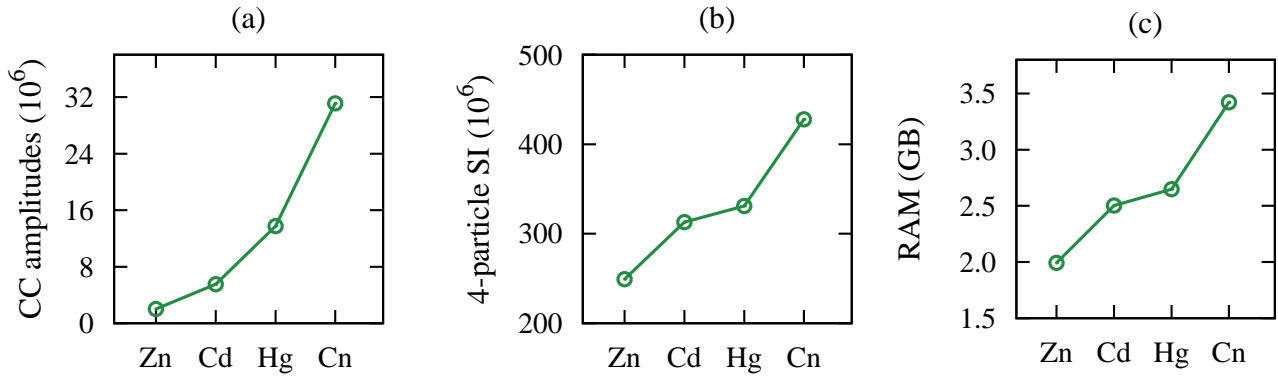}
  \caption{(a) Number of cluster amplitudes, (b) number of 4-particle 
           Slater integrals and (c) memory required to store 4-particle 
           Slater integrals, as a function of $Z$ for group-12 elements.}
  \label{cc_amp}
\end{figure*}


\subsection{Computational challenges with SHEs}

  The calculation of $\alpha$ for SHEs is a computationally challenging task.
This is due to the large number of core electrons and the need for larger 
basis size to obtain converged properties results. The latter pose three 
main hurdles in the calculations. First, the number of cluster amplitudes
is very large, and solving the cluster equations require long compute times. 
To give an example, as shown in Fig. \ref{cc_amp}, in the case of Cn using 
a converge basis of 200 orbitals  leads to more than 31 millions cluster
amplitudes. This is about 2.3 times larger than the lighter atom Rn.  
Second, convergence of $\alpha$ with basis size is slow. This is in stark
contrast to the convergence trends of $\alpha$ for lighter atoms and ions
reported in our previous works \citep{chattopadhyay-14, chattopadhyay-15,
ravi-20}. For the lighter atoms and ions convergence is achieved with a 
basis of 160 or less orbitals. However, for SHEs  convergence of $\alpha$
requires $\approx$ 200 orbitals.  And third, storing the two-electron 
integrals for efficient computation require of large memory. For instance, 
the number of {\em 4-particles} two-electron integrals in the case of Cn 
is more than 427 millions. This is about 1.3 times larger than the case of Rn. 
Moreover, in general parallelization, solving the cluster equations require
storing the same set of integrals are stored across all nodes. This leads 
to replication of data across compute nodes and puts severe restrictions 
on basis size in the PRCC calculations. To mitigate this problem, we have 
implemented a {\em memory-parallel-storage} algorithm \cite{mani-17} which 
avoids the storage replication of the integrals across different nodes. This 
allows efficient memory usage and use large orbital basis in the PRCC 
computations. The inclusion of perturbative triples to the computation of 
$\alpha$ enhances the computational complexity further. This is due to the 
evaluation of numerous additional polarizability diagrams arising from the
perturbative triples. To illustrate the compute time, the computation of 
$\alpha$ for Cn using a basis of 200 orbitals without triples takes 120 hrs 
with 144 threads. Whereas, with partial triples included, it requires 280 hrs
with 200 threads. Thus the runtime more than doubled.

\section{Results and Discussion}

\subsection{Convergence of $\alpha$ and correlation energy}  

\begin{figure*}
 \includegraphics[height=17cm, angle=-90]{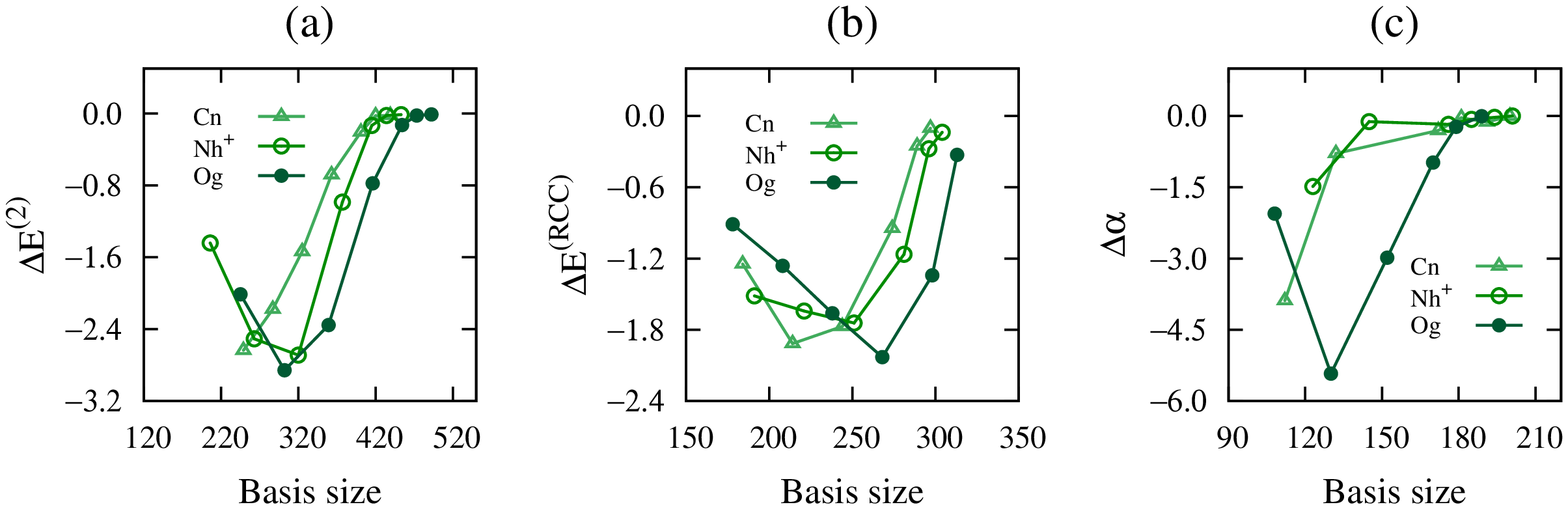}
	\caption{ Convergence of second-order correlation energy (panel (a)), 
	the RCC energy (panel (b)) and $\alpha$ (panel (c)) as function of 
	the basis size.}
\label{fig_conv}
\end{figure*}

  Since the GTO basis are mathematically incomplete \cite{grant-06}, it is 
essential to check the convergence of polarizability and correlation
energy with basis size. The convergence trends of $\alpha$ and electron 
correlation energy are shown in the Fig. \ref{fig_conv}. In these calculations 
we have used the Dirac-Coulomb Hamiltonian as it is computationally less 
expensive than using the DCB Hamiltonian. To determine the converged basis set 
we start with a moderate basis size and add orbitals in each symmetry 
systematically. This is continued till the change in $\alpha$ and correlation 
energy is $\leqslant 10^{-3}$ in their respective units. For example, 
as discernible from the Table \ref{con-alpha} in Appendix, the change in 
$\alpha$ for Cn is $4.0\times10^{-3}$ a.u. when the basis set is augmented 
from 191 to 200. So, to optimize the compute time, we consider the basis set 
with 191 orbitals as the optimal one for $\alpha$. Once the optimal basis 
set is chosen, for further computations we incorporate the Breit interaction 
and QED corrections. As discernible from the Figs. \ref{fig_conv}(a) and (b), 
one key observation is, the correlation energies converge with much 
larger bases than $\alpha$. For example, for Cn, the 
converged second-order energy is obtained with the basis size of 439 
($31s27p26d24f22g21h21i21j21k21l$) 
orbitals. A similar trend is also observed for other two SHEs and all 
lighter homologs considered in this work.


\subsection{Correlation energy}

The electron correlation energy in RCC is expressed as
\begin{equation}
	\Delta E = \langle\Phi_0|\bar{H}_{\rm N}|\Phi_0\rangle,
\end{equation}
where ${\bar H}_{\rm N}, = e^{-T^{(0)}} H_{\rm N} e^{T^{(0)}},$ is the
similarity transformed Hamiltonian. In Table \ref{corr-ene}, we list
$\Delta E$ for SHEs and three lighter elements in each group. Since the 
correlation energies converge with very large basis sizes, it is therefore not 
practical to use such a large basis in the RCC calculations due to several 
computational limitations. Some of the limitations are as mentioned in the 
previous section. To mitigate this, and to account for correlation energy from 
the virtuals not included in the RCC calculations, we resort to the second-order 
MBPT method. The RCC results for $\Delta E$ listed in Table \ref{corr-ene} are 
calculated using the expression
\begin{equation}
	\Delta E_{\rm RCC} \approx \Delta E^{\rm nconv}_{\rm RCC} +
	\left(\Delta E^{\rm conv, 2}_{\rm MBPT} - 
	\Delta E^{\rm nconv, 2}_{\rm MBPT} \right),
\end{equation}
where $\Delta E^{\rm nconv}_{\rm RCC}$ is the correlation energy computed
using RCC with orbitals up to $j$-symmetry. And, 
$\Delta E^{\rm nconv, 2}_{\rm MBPT}$ and
$\Delta E^{\rm conv, 2}_{\rm MBPT}$ are the second-order energies 
calculated using RCC basis and a converged basis which includes orbitals up 
to $l$-symmetry, respectively.
\begin{table}[h!]
\caption{Electron correlation and total energies in atomic units for
    group-12, group-13 and group-18 elements. Listed RCC energies also
    include the contributions from the Breit and QED corrections.}
   \label{corr-ene}
   \begin{center}
   \begin{ruledtabular}
   \begin{tabular}{ccccccc}
  &\multicolumn{2}{c}{Basis} & \multicolumn{2}{c}{$\Delta E_{\rm DC}$} &
  $E_{\rm total}$ & $\Delta E_{\rm Others}$  \\
                    \hline
   & MBPT & RCC & MBPT & RCC & &  \\
                    \hline
                    \multicolumn{7}{c}{Group-12}  \\
                    \hline
  Zn & $336$ & $206$ & $-1.6769$ & $-1.5690$ & $-1796.1812$ &$-1.6975^a$,  \\
           &       &       &  &                  &       &   $-1.6611^d$   \\
           &       &       &  &                  &       &   $-1.6206^f$   \\
  Cd & $461$ & $223$ & $-2.7278$ & $-2.6216$ & $-5595.9393$ &$-2.7253^b$,  \\
           &       &       &  &                  &       &   $-2.6540^d$   \\
           &       &       &  &                  &       &   $-2.6500^f$   \\
  Hg & $413$ & $227$ & $-5.4681$ & $-5.1164$ & $-19653.9388$&$-5.4508^b$,  \\
           &       &       &  &                  &       &   $-5.2895^d$   \\
           &       &       &  &                  &       &   $-5.1760^f$   \\
  Cn & $439$ & $289$ & $-8.4393$ & $-7.7981$  &  $-47335.9752$         &   \\
                    \multicolumn{7}{c}{Group-13}  \\
                    \hline
  Ga$^+$ & $411$ & $227$ & $-1.669$ & $-1.58077$ & $-1943.9435$  &         \\
                           \\
  In$^+$ & $447$ & $235$ & $-2.744$ & $-2.64147$ & $-5882.8838$  &         \\
                           \\
  Tl$^+$ & $409$ & $220$ & $-5.499$ & $-5.15772$ & $-20279.7851$ &         \\
                           \\
  Nh$^+$ & $453$ & $304$ & $-8.4743$ &  $7.8439$ & $-48517.6211$ &         \\
                    \multicolumn{7}{c}{Group-18}  \\
                    \hline
 Kr     & $413$ & $255$ & $-1.8532$ & $-1.7900$ & $-2790.3898$  &$-1.8907^c$, \\
           &       &       &  &                  &       & $-1.8468^d$        \\
           &       &       &  &                  &       & $-1.8466^e$        \\
           &       &       &  &                  &       & $-1.8496^f$        \\
  Xe     & $419$ & $255$ & $-3.0314$ & $-2.9075$ & $-7448.6635$  &$-3.0877^c$,\\
           &       &       &  &                  &       & $-2.9587^d$        \\
           &       &       &  &                  &       & $-2.9979^e$        \\
           &       &       &  &                  &       & $-3.0002^f$        \\
  Rn       & $372$ & $245$ & $-5.6195$ & $-5.2945$ &$-23601.3243$&$-5.7738^c$,\\
           &       &       &  &                  &       & $-5.5874^d$        \\
           &       &       &  &                  &       & $-5.5250^f$        \\
  Og       & $492$ & $313$ & $-8.9109$ & $-8.3047$  & $-54815.0764$      &
   \end{tabular}
   \end{ruledtabular}
   \end{center}
   \begin{tabbing}
  $^{\rm a}$Ref.\cite{flores-93a}[MP2],  \\
  $^{\rm b}$Ref.\cite{flores-94}[MP2],   \\
  $^{\rm c}$Ref.\cite{flores-93b}[MP2],  \\
  $^{\rm d}$Ref.\cite{ishikawa-94}[MP2], \\
  $^{\rm e}$Ref.\cite{mani-09}[RCC],     \\
  $^{\rm f}$Ref.\cite{mccarthy-11}[RCC],     \\
\end{tabbing}
\end{table}

\begin{figure*}
 \includegraphics[height=17cm, angle=-90]{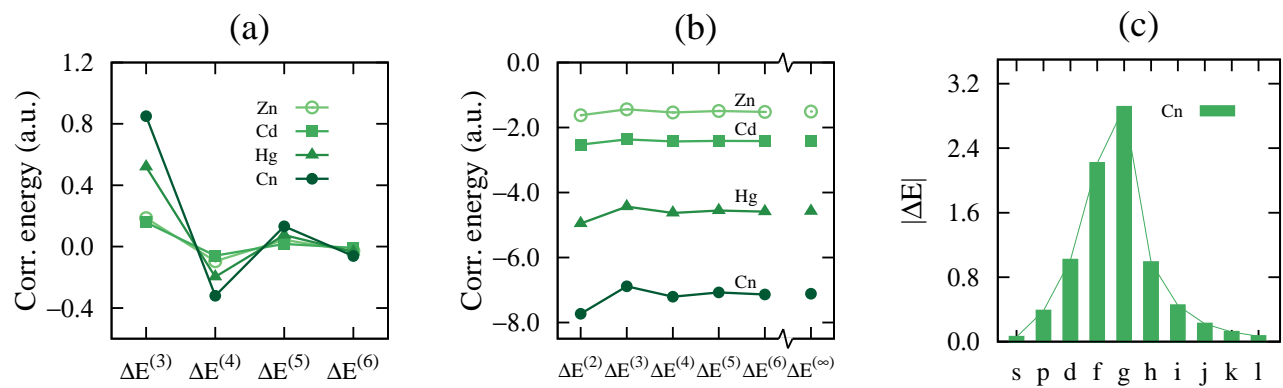}
	\caption{(a) Third, fourth, fifth and sixth-order correlation 
	energies, (b) cumulative correlation energy and (c) contribution
	to correlation energy from orbitals of different symmetries. 
	$\Delta E^{(\infty)}$ in panel (b) represents the infinite-order 
	correlation energy and is equivalent to RCC energy.}
 \label{fig_corr}
\end{figure*}

For all the elements listed in Table \ref{corr-ene}, we observe an important
trend in the correlation energy. The RCC energy is smaller in magnitude than 
the second-order energy. 
A similar trend is also observed in the previous 
work \cite{mccarthy-11}. From our calculations we find that the reason for 
this is the cancellations from the higher order corrections embedded in RCC. 
As discernible from the Fig. \ref{fig_corr}(a) for group-12 elements as an 
example, contributions from the third and fifth order corrections are 
positive. And, as a result, correlation energy oscillates initially before 
converging to the RCC value. This can be observed from the Fig. \ref{fig_corr}(b) 
where we have shown the cumulative contribution.
Comparing our results with previous calculations, to the best of our knowledge, 
there are no results for SHEs. For lighter elements, our RCC energy agrees well
with the previous RCC results \cite{mccarthy-11,mani-09}. The small difference
in the energies, however, could be attributed to the corrections from the
Breit and QED included in the present work and the difference 
in the basis employed.
For second-order energies, there are four results from the previous 
calculations \cite{flores-93a,flores-94,flores-93b,ishikawa-94}. And, our 
results match very well with them for all the elements. 

Examining the contributions from different symmetries of virtual orbitals, we 
find that all three SHEs show a similar trend. This is not surprising as 
all three are closed-shell systems. 
As discernible from the histograms in Fig. \ref{fig_corr}, contribution to 
correlation energy increases initially as a function of orbital symmetry and 
then decreases. 
The first two dominant changes, $\approx$ 35\% and 26\% of the total 
correlation energy, are from the $g$ and $f$ orbitals. The next two are from 
the $d$ and $h$-symmetries, they contribute $\approx$ 12\% each to all the 
SHEs. The contribution from the virtuals with $l$-symmetry is about 0.8\%. This
non negligible contribution from $l$-symmetry orbitals indicates that the 
inclusion of the orbitals from higher symmetries are important to obtained 
the accurate energies for SHEs.


\subsection{Polarizability}

The values of $\alpha$ with different methods subsumed in the PRCC theory 
are listed in the Table \ref{final-alpha}. The Dirac-Fock (DF) contribution 
is computed using Eq. (\ref{dbar}) with $\mathbf{T}^{(1)}$ and 
$\bar{\mathbf{D}}$ replaced by the bare dipole operator $\mathbf{D}$, and 
is expected to have the dominant contribution. The PRCC values are the
converged values from Table \ref{con-alpha}, calculated using the DC 
Hamiltonian with basis up to $h$ symmetry.  The values listed as {\em 
Estimated} include the estimated contribution from the orbitals of $i$, $j$ 
and $k$ symmetries. For this, we use a basis set of moderate size from the
Table \ref{con-alpha} and then, augment it with orbitals from $i$, $j$ and 
$k$ symmetries to calculate percentage contribution, this is added to the 
PRCC value. To the best of our knowledge, there are no experimental data on
$\alpha$ for SHEs considered in the present work. However, to understand 
the trend of electron correlation effects, we compare our results with the 
previous theoretical results. One important and crucial difference between 
the previous studies and the present work is the absence of QED corrections 
in the previous works. Though the Breit interaction is included in the 
previous work \cite{seth-97} for Cn, the contribution is not given 
explicitly. These corrections are, however, important to obtain the accurate 
and reliable values of $\alpha$ for SHEs. From our calculations, we find 
that the combine Breit+QED contributions are $\approx$ 0.5$\%$, 0.4\% and 
0.6\%, respectively, for Cn, Nh$^+$ and Og. Considering the important 
prospects associated with accurate data on $\alpha$ for SHEs, these are 
significant contributions and can not be neglected.
\begin{table}
	\caption{Final value of $\alpha$ (a. u.) from PRCC calculation compared 
	with other theoretical data in the literature.}
  \label{final-alpha}
  \begin{ruledtabular}
  \begin{tabular}{clcc}
      Element & \multicolumn{2}{c}{\textrm{Present work}} & {\textrm{Other cal.}}\\
            \hline
           & {\textrm{Method}}& $\alpha$ &  \\
            \hline
      Cn   & DF   & $35.234$ &  $25.82^{\rm a}$, $28.68^{\rm b}$,    \\
           & PRCC & $26.944$ &  $27.40^{\rm d}$, $28\pm4^{\rm c}$ \\
           & PRCC(T)& $27.457$ &   \\
           & PRCC(T)+Breit & $27.537$ &       \\
           & PRCC(T)+Breit+QED & $27.588$ \\
           & Estimated & $27.442$  &  \\
	   & Recommended &$27.44(88)$  &  \\
            \hline
      Nh$^+$ & DF   & $23.182$       \\
           &   PRCC & $17.056$       \\
           & PRCC(T)& $17.063$        \\
           & PRCC(T)+Breit & $17.100$ \\
           & PRCC(T)+Breit+QED & $17.135$ \\
             & Estimated & $17.123$  &  \\
	     & Recommended & $17.12(55)$  &  \\
            \hline
      Og & DF   & $56.197$ &  $52.43^{\rm b}$, $46.33^{\rm e}$,  \\
           & PRCC     & $55.941$   &  $57.98^{\rm f}$, $57\pm3^{\rm c}$  \\
	   & PRCC(T)& $56.203$ &  \\
           & PRCC(T)+Breit & $56.250$ &      \\
	   & PRCC(T)+Breit+QED & $56.545$ &  \\
	     & Estimated & $56.536$  & \\
	     & Recommended & $56.54(181)$  &  \\
        \end{tabular}
        \end{ruledtabular}
\begin{tabbing}
  $^{\rm a}$Ref.\cite{seth-97}[CCSD(T)],  \\
  $^{\rm b}$Ref.\cite{nash-05}[CCSD(T)],  \\
  $^{\rm c}$Ref.\cite{dzuba-16}[RRPA],  \\
  $^{\rm d}$Ref.\cite{pershina-08a}[DC-CCSD(T)],  \\
  $^{\rm e}$Ref.\cite{pershina-08b}[R, DC-CCSD(T)],  \\
  $^{\rm f}$Ref.\cite{jerabek-18}[R, Dirac$+$Gaunt, CCSD(T)],  \\
\end{tabbing}
\end{table}
\begin{table}
    \caption{Contributions to $\alpha$ (in a.u.) from different terms in 
     PRCC theory.}
    \label{termwise}
    \begin{center}
    \begin{ruledtabular}
    \begin{tabular}{lccr}
        Terms + h.c. & \multicolumn{1}{r}{$\rm{Cn}$}
        & \multicolumn{1}{r}{$\rm{Nh^+}$}
        & \multicolumn{1}{r}{$\rm{Og}$}  \\
        \hline
        $\mathbf{T}_1^{(1)\dagger}\mathbf{D} $            & $34.5267$ & $21.3767$  & $68.9516$  \\
        $\mathbf{T}_1{^{(1)\dagger}}\mathbf{D}T_2^{(0)} $ & $-3.0095$ & $-1.7444$  & $-4.0179$  \\
        $\mathbf{T}_2{^{(1)\dagger}}\mathbf{D}T_2^{(0)} $ & $ 1.7485$ & $ 0.7622$  & $ 3.1900$  \\
        $\mathbf{T}_1{^{(1)\dagger}}\mathbf{D}T_1^{(0)} $ & $-0.0389$ & $-0.0809$  & $-1.5258$  \\
        $\mathbf{T}_2{^{(1)\dagger}}\mathbf{D}T_1^{(0)} $ & $-0.1087$ & $-0.0247$  & $ 0.2319$  \\
        Normalization                                     & $ 1.2292$ & $ 1.1898$  & $ 1.1946$  \\
        Total                                             & $26.9435$ & $17.0524$  & $55.9432$  \\
    \end{tabular}
    \end{ruledtabular}
    \end{center}
\end{table}

For Cn three of the previous studies, similar to the present work, are  using 
CCSD(T). There are, however, important differences in terms of the basis 
used in these calculations. And, this could account for the difference in 
the values of $\alpha$ reported in these works. In Ref. \cite{seth-97},
a relativistic basis with $11s 8p 8d 4f$ orbitals optimized using 
pseudopotential Hartree-Fock energy is used and reports the smallest
value 25.82. The other CCSD(T) result 27.40 from the Ref. \cite{pershina-08a} 
is using a relatively larger basis of $26s24p18d13f5g2h$.  In terms of 
methodology and basis, Ref. \cite{pershina-08a} is the closest to the 
present work. Our recommended value $27.44(88)$ is close this.
The other CCSD(T) result of 28.68 is from the Ref. \cite{nash-05}, which is
obtained using an uncontracted Cartesian basis. The is the highest theoretical
value reported in the literature. The other value $28\pm4$ is obtained
using the RRPA \cite{dzuba-16}. And, this is close to our result. This is
is to be expected  as both RRPA and PRCC account core-polarization, which is 
the dominant contribution to $\alpha$ after the DF. For the triples 
contribution to $\alpha$, there is no clear trend in the previous RCC results. 
In Ref. \cite{seth-97} and \cite{pershina-08a} the contribution from 
triples reported to as $-0.08$ and $-0.07$\% of the CCSD value, respectively, 
and decrease the value of $\alpha$. However, a positive contribution 
of $\approx$ 0.25\% is reported in Ref. \cite{nash-05}. In the present 
work, we obtain a positive contribution of $\approx$ 1.9\%, which increases 
the value of $\alpha$ further.

For Nh$^+$, there are no previous theoretical results. The present work 
reports the first theoretical result of $\alpha$, using
PRCC theory. As we observed from the Table \ref{final-alpha}, though 
it has the same electronic structure as Cn, the value of $\alpha$ is 
smaller. This is attributed to the relativistic contraction of $7s_{1/2}$ 
orbital due to increased nuclear potential. Like the case of Cn, the 
inclusion of partial triples increases the value of $\alpha$ further.

For Og, there are three previous results based on calculations using CCSD(T). 
Though the same methods are used, there is a large difference in the 
values of $\alpha$ reported in these works. For instance, the CCSD(T) value
57.98 reported in Ref. \cite{jerabek-18} is $\approx$ 25\% larger than the
result in Ref. \cite{pershina-08b}. The reason for this could be attributed 
to the different types of basis used. In Ref. \cite{pershina-08b} the
computations used the Faegri basis with $26s24p18d13f5g2h$ orbitals, 
however, in Ref. \cite{jerabek-18} an uncontracted relativistic 
quadrupole-zeta basis is used. The other CCSD(T) result 52.43 from 
Ref. \cite{nash-05} lies between the other two results. Our recommended
value $56.54(181)$ is closer to the RRPA value $57(3)$ 
from Ref. \cite{dzuba-16} and CCSD(T) value, 57.98, from 
Ref. \cite{jerabek-18}. As mentioned in the case of Cn, this is due to 
core-polarization effect accounted to all orders in both CCSD and RRPA.
The obtained contribution from partial triples 0.47\% is consistent with 
the contribution 0.66\% reported in Ref. \cite{pershina-08b}.


\section{Electron correlation, Breit and QED corrections}  

  In this section we analyze and present the trends of electron correlation 
effects from the residual Coulomb interaction, Breit interaction and 
QED corrections to $\alpha$ as function of $Z$.

\subsection{Residual Coulomb interaction}

\begin{table}
\caption{Five leading contribution to $\{ \bf{ T_1^{(1)\dagger} D} + H.c.\}$ 
    (in a.u.) for $\alpha$ from core orbitals. This includes the DF and 
    core-polarization contributions.}
\label{t1d}
\begin{center}
\begin{ruledtabular}
\begin{tabular}{ccc}
   $\rm{Cn}$  &  $\rm{Nh^+}$  &  $\rm{Og}$  \\
    \hline
    $17.468(7s_{1/2})$ & $11.440(7s_{1/2})$ & $65.874(7p_{3/2})$  \\
    $12.374(6d_{5/2})$ & $ 6.534(6d_{5/2})$ & $ 2.392(7p_{1/2})$  \\
    $ 4.776(6d_{3/2})$ & $ 3.308(6d_{3/2})$ & $ 0.562(6d_{5/2})$  \\
    $ 0.052(6p_{3/2})$ & $ 0.224(6p_{3/2})$ & $ 0.264(6d_{3/2})$  \\
    $ 0.015(5f_{7/2})$ & $ 0.002(5f_{7/2})$ & $ 0.042(7s_{1/2})$  \\
\end{tabular}
\end{ruledtabular}
\end{center}
\end{table}

\begin{table}
\caption{Five leading contributions to NLO term 
 $\{\bf{ T_1^{(1)\dagger}D \ T_2^{(0)} + H.c.} \}$ (in a.u.) for 
 $\alpha$ from core-core orbital pairs. This includes the pair-correlation 
 contributions.}
\label{t1dt2}
\begin{center}
\begin{ruledtabular}
\begin{tabular}{cc}
    $\rm{Cn}$  &  $\rm{Nh^+}$ \\
    \hline
    $-0.618(7s_{1/2},6d_{5/2})$&$-0.420(7s_{1/2},6d_{5/2})$ \\  
    $-0.479(7s_{1/2},7s_{1/2})$&$-0.243(7s_{1/2},7s_{1/2})$ \\  
    $-0.405(6d_{5/2},6d_{5/2})$&$-0.235(7s_{1/2},6d_{3/2})$ \\  
    $-0.337(6d_{5/2},7s_{1/2})$&$-0.204(6d_{5/2},6d_{5/2})$ \\  
    $-0.327(7s_{1/2},6d_{3/2})$&$-0.171(6d_{5/2},7s_{1/2})$ \\ 
    $\rm{Og}$   \\
    $-2.805(7p_{3/2},7p_{3/2})$ \\
    $-0.446(7p_{3/2},7p_{1/2})$ \\
    $-0.365(7p_{3/2},6d_{5/2})$ \\
    $-0.167(7p_{3/2},6d_{3/2})$ \\
    $-0.114(7p_{1/2},7p_{3/2})$  \\
\end{tabular}
\end{ruledtabular}
\end{center}
\end{table}
\begin{figure*}
\includegraphics[height=17cm, angle=-90]{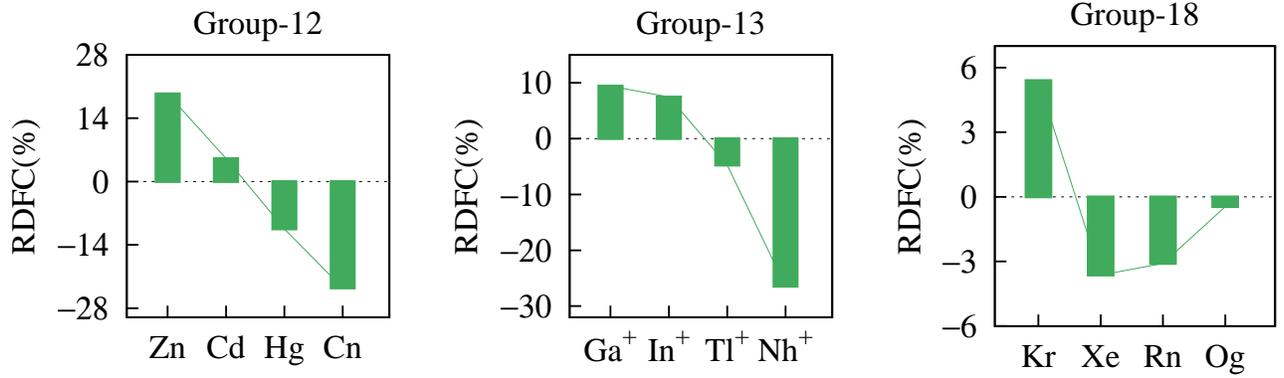}
	\caption{In percentage, the { \em relative-DF-contribution } for 
	group-12, group-13 and group-18 elements.} 
\label{fig_rdfc}
\end{figure*}

  To assess the correlation effects from residual Coulomb interaction we 
define {\em relative-DF-contribution} (RDFC) as 
$$
  {\rm RDFC} = \frac{\alpha_{\rm PRCC} - \alpha_{\rm DF}}{\alpha_{\rm DF}},
$$
and plot this for each of the groups in Fig. \ref{fig_rdfc} for all the
four elements. As observed from the figure, we obtain similar trends
for the group-12 and group-13 elements. 
For these groups, the RDFC is positive initially and then changes to negative. 
The reason for this is the drastic change in the nature of the core polarization 
contribution as function of Z, due to different screening of nuclear potential. 
The core polarization contribution is positive for first two elements and negative 
for the last two. This negative contribution reduces the PRCC value to lower 
than the DF value.
A similar trend is also reported in the previous 
works \cite{seth-97,nash-05,pershina-08a} where the DF value for Cn is 
higher than the CCSD value. 
For the group-18 elements, the RDFC shows a slightly different trend. Except 
for Kr, it is negative for all the remaining elements. In addition, the 
magnitude decreases from Xe to Og. This could be attributed to the negative 
and decreasing core polarization contributions from Xe to Og.
Our higher DF value, 56.20, for Og is consistent with the previous results in 
Refs. \cite{nash-05,pershina-08b} in which the reported DF values of 54.46 
and 50.01, respectively, are larger than the CCSD values. The difference 
in the DF values could be due to the different basis used in these 
calculations, which also led to the different $\alpha$ values.  
\begin{figure*}
\includegraphics[height=17cm, angle=-90]{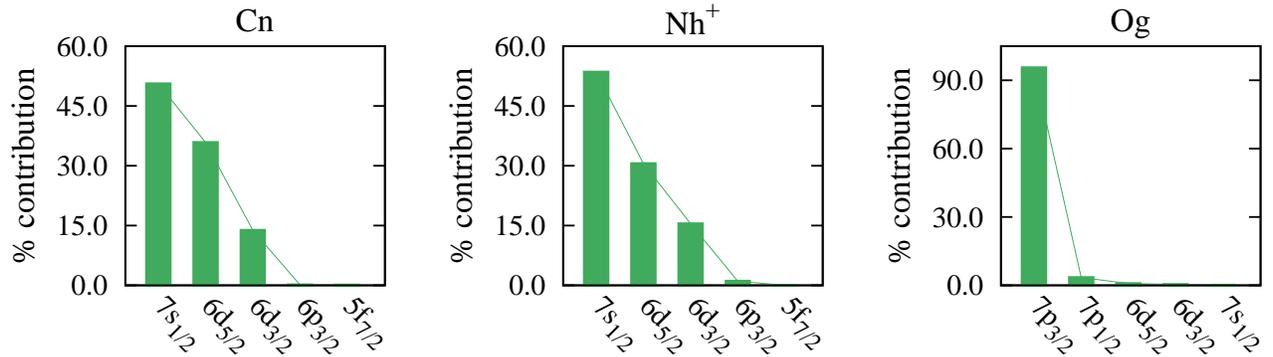}
\caption{Five largest percentage contribution from core orbitals to LO term 
	$\{\mathbf{T}_1^{(1)\dagger}\mathbf{D} + {\rm h.c.}\}$.}
\label{fig_t1d}
\end{figure*}

\begin{figure*}
\includegraphics[height=17cm, angle=-90]{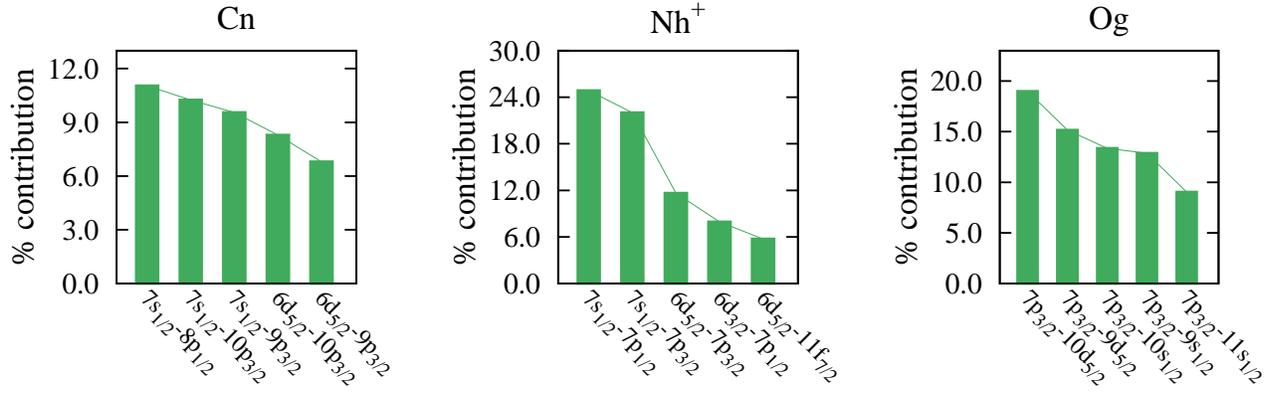}
 \caption{In percentage, five dominant dipolar mixing of cores with 
	 virtuals.}
\label{fig_t1d2}
\end{figure*}

\begin{figure*}
\includegraphics[height=17cm, angle=-90]{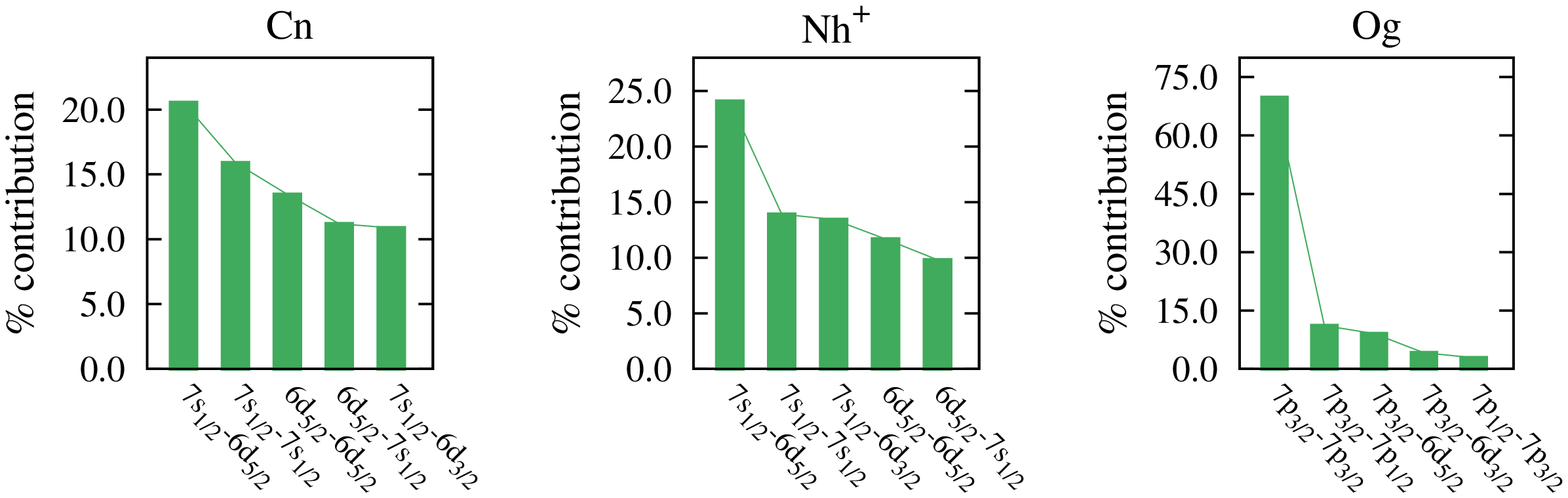}
 \caption{Five largest percentage contribution from core-core orbital pairs 
	  to NLO term.}
\label{fig_t1dt2}
\end{figure*}

  To gain further insights on the electron correlations effects subsumed 
in the PRCC theory, we examine the contributions from different terms.
And, these are listed in the Table \ref{termwise}. As seen from
the table, for all the SHEs, the LO contribution is from the term 
$\{\mathbf{T}_1^{(1)\dagger}\mathbf{D} + {\rm h.c.}\}$. This is to be 
expected, as it subsumes the contributions from DF and RPA. The contributions 
are larger than PRCC by $\approx$ 28\%, 25\% and 23\% for Cn, Nh$^+$ and Og,
respectively. The contribution from the NLO term  
$\{\mathbf{T}_1{^{(1)\dagger}}\mathbf{D}T_2^{(0)}\}$ is small and
opposite in phase to the LO term. It accounts for $\approx$ -11\%,
-10\% and -7\% of the PRCC value for Cn, Nh$^+$ and Og, respectively.
The next to NLO (NNLO) term is 
$\mathbf{T}_2{^{(1)\dagger}}\mathbf{D}T_2^{(0)}$ and contributes 
$\approx$ 6\%, 4\% and 6\% of the PRCC value. The contributions from the 
other terms are small, and the reason is the smaller magnitude of the 
$T_1^{(0)}$ CC operators.

  To examine in more detail, we assess the contributions from the 
core-polarization and pair-correlation effects. For the core-polarization, 
we identify five dominant contributions to the LO term and these are listed 
in the Table \ref{t1d}. Since the Cn and Nh$^+$ have the same ground state
electronic configuration, both show similar correlation trends. For both, the 
most dominant contribution is from the valence orbital $7s_{1/2}$ and this
is due to its larger radial extent. As shown in the Fig. \ref{fig_t1d}, 
the contribution from $7s_{1/2}$ is $\approx$ 50\% and 53\% of the LO value
for Cn and Nh$^+$, respectively.  For Cn, we find that more than 60\% of the
$7s_{1/2}$ contribution arise from $\mathbf{T}_1{^{(1)}}$ involving
the $8p_{1/2}$, $10p_{3/2}$ and $9p_{3/2}$ orbitals. Whereas for Nh$^+$, 
the $7p_{1/2}$ and $7p_{3/2}$ together contribute more than 87\% of the total 
contribution. The next two important contributions are from the 
core orbitals $6d_{5/2}$ and $6d_{3/2}$. The contribution from the former is
almost double of the latter. In particular, for the Cn and Nh$^+$ the 
contributions from $6d_{5/2}$ is 35\% and 30\%, respectively. Whereas, 
the contribution from $6d_{3/2}$ is $\approx$ 14\% and 15\%, respectively. 
The larger contribution from $6d_{5/2}$ could be attributed to the strong 
dipolar mixing with $10p_{3/2}$ and $9p_{3/2}$ for Cn, and $7p_{3/2}$ and 
$11f_{7/2}$ for Nh$^+$ (see the Fig. \ref{fig_t1d2}).

  For Og, compare to Cn and Nh$^+$, we observe a different trend of 
core-polarization effect. More than 95\% of the contribution from the LO 
term arises from valence orbital $7p_{3/2}$. The other valence and core orbitals 
contribute less than 5\% and $7p_{1/2}$ contributes only $\approx$ 3\% of 
the LO term. The reason for this could be the larger radial extent of 
the $7p_{3/2}$ orbital as $7p_{1/2}$ orbital contracts due to relativistic
effects. The five dominant contributions arise from the dipolar mixing of 
$7p_{3/2}$ with $10d_{5/2}$, $9d_{5/2}$, $10s_{1/2}$, $9s_{1/2}$ and 
$11s_{1/2}$ orbitals. These orbitals together contribute $\approx$ 73\% of 
the total contribution (see the  Fig. \ref{fig_t1d2}).

  To assess the contribution from pair-correlation effects we consider the 
NLO term and identify the dominant contributions to it. These are listed 
in the Table \ref{t1dt2} in terms of the pairs of core orbitals and these
correspond to the $T_2^{(0)}$ with dominant contributions. This is an 
appropriate approach as the most dominant term involving doubly excited 
cluster operator is the NLO term. For better illustration the percentage 
contribution to those listed in Table \ref{t1dt2} are  plotted in the 
Fig. \ref{fig_t1dt2}. For both Cn and Nh$^+$, the first two dominant 
contributions are from the $(7s_{1/2},6d_{5/2})$ and $(7s_{1/2},7s_{1/2})$ 
core-orbital pairs. In percentage, these are $\approx$ -20\% and -16\% 
for Cn, whereas $\approx$ -24\% and -14\% for Nh$^+$. 
Though the next three contributions are from the same 
core-orbital pairs, $(6d_{5/2},6d_{5/2})$, $(6d_{5/2},7s_{1/2})$ and 
$(7s_{1/2},6d_{3/2})$, in both the elements, there are differences in terms 
of the order in which they contribute. Like in the core-polarization effect, 
we observe a different trend for Og. About 70\% of the total contribution is 
from only the $(7p_{3/2},7p_{3/2})$ orbital pair.

\subsection{Breit and QED corrections}

\begin{table}
    \caption{Contributions to $\alpha$ from Breit interaction, vacuum
        polarization and the self-energy corrections in atomic units.}
    \label{breit_qed}
    \begin{center}
    \begin{ruledtabular}
    \begin{tabular}{llccc}
        Elements & $Z$ & Breit int. & Self-ene.   & Vacuum-pol.\\
                       \hline
                       \multicolumn{3}{r}{Group-12} \\
                       \hline
           Zn  &  $30$ &   $-0.0928$  & $0.0221$ &  $-0.0038$  \\
           Cd  &  $48$ &   $-0.0953$  & $0.0648$ &  $-0.0159$  \\
           Hg  &  $80$ &   $ 0.0519$  & $0.0933$ &  $-0.0358$  \\
           Cn  & $112$ &   $ 0.0802$  & $0.1072$ &  $-0.0557$  \\
                       \multicolumn{3}{r}{Group-13} \\
                       \hline
           Ga$^+$  &  $31$ &   $-0.3006$  & $0.0090$ &  $-0.0018$  \\
           In$^+$  &  $49$ &   $-0.3647$  & $0.0249$ &  $-0.0070$  \\
           Tl$^+$  &  $81$ &   $-0.1283$  & $0.0526$ &  $-0.0274$  \\
           Nh$^+$  & $113$ &   $ 0.0366$  & $0.0794$ &  $-0.0440$  \\
                       \multicolumn{3}{r}{Group-18} \\
                       \hline
           Kr  &  $36$ &   $0.0179$  & $0.0011$ &  $0.0009$  \\
           Xe  &  $54$ &   $0.0213$  & $0.0042$ &  $0.0031$  \\
           Rn  &  $86$ &   $0.0226$  & $0.0239$ &  $0.0181$  \\
           Og  & $118$ &   $0.0472$  & $0.1162$ &  $0.1769$  \\
    \end{tabular}
    \end{ruledtabular}
    \end{center}
\end{table}

\begin{figure*}
 \includegraphics[height=17cm, angle=-90]{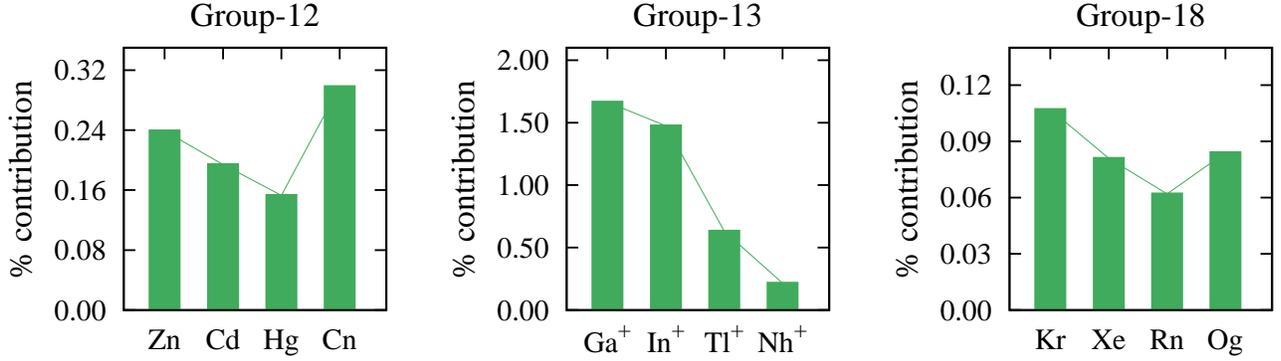}
 \caption{Percentage contribution from Breit interaction to group-12, 
          group-13 and group-18 elements.}
 \label{fig_breit}
\end{figure*}

\begin{figure*}
 \includegraphics[height=17cm, angle=-90]{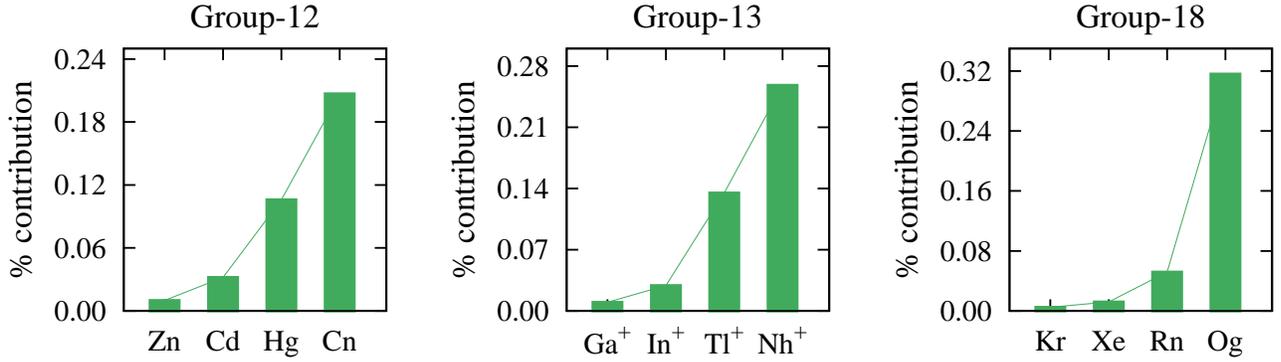}
 \caption{Percentage contribution from vacuum polarization to group-12, group-13
          and group-18 elements.}
 \label{fig_vacu}
\end{figure*}

\begin{figure*}
 \includegraphics[height=17cm, angle=-90]{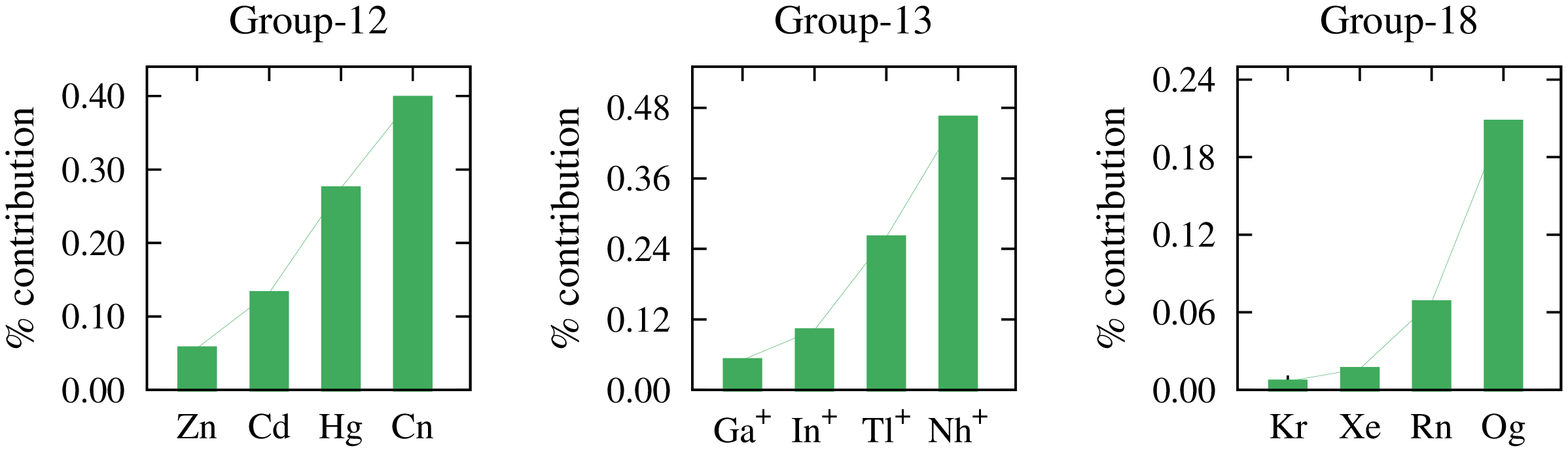}
 \caption{Percentage contribution from self-energy correction to group-12, 
          group-13 and group-18 elements.}
 \label{fig_self}
\end{figure*}

To analyze the trend of correlation effects arising from the Breit 
interaction, vacuum polarization and the self-energy corrections as function 
of $Z$, we separate the contributions from these interactions. And, these 
are listed in the Table \ref{breit_qed}. In addition, for comparison and
to show the trends in the group, the percentage contributions from the 
corresponding groups in the periodic table of the SHEs are shown in 
Figs. \ref{fig_breit}, \ref{fig_vacu} and \ref{fig_self}, respectively. 
For the Breit 
interaction, as we see from the Fig. \ref{fig_breit}, except for the Cn and 
Og, we observe a trend of decreasing contributions with increasing $Z$ within 
the groups.
One common trend we observe in the contributions to SHEs is that, all are in 
the same phase as PRCC, and hence increase the value of $\alpha$. For lighter 
elements, however, we get a mixed phase for contributions.

For the corrections from the vacuum polarization and self energy, from the 
Figs. \ref{fig_vacu} and \ref{fig_self} we see 
that, the contribution from both the vacuum polarization and self energy 
increases with $Z$ for all the three groups. This is as expected. For 
the vacuum polarization, the effect is larger due to higher nuclear charge $Z$.  
And, for the self energy, the correction depends on the energy of the orbital, 
which again depends on the nuclear charge. In terms of the phase of the 
contributions from vacuum polarization, these are in opposite to PRCC value 
for all the elements of group-12 and group-13, and hence lowers the 
value of $\alpha$. For group-18, however, we observe the contributions 
of the same phase as PRCC.  In terms of magnitude, the contributions are 
$\approx$ 0.21$\%$, 0.26$\%$ and 0.31$\%$ of the PRCC value for Cn, Nh$^+$ 
and Og, respectively. For the self energy, one prominent feature of the 
contributions we observe is that it is positive for all elements in all the
three groups, and therefore increase the value of $\alpha$. The contributions 
in the case of Cn, Nh$^+$ and Og are $\approx$ 0.4\%, 0.5\% and 0.2\% of 
the PRCC value, respectively.

\subsection{Theoretical uncertainty} 

  In this section we discuss the theoretical uncertainty associated with 
our results for $\alpha$. 
For this, we have identified four different sources which
can contribute. The first source of uncertainty is the truncation of 
the basis set in our calculations. The recommended values of $\alpha$ in 
Table \ref{final-alpha} are obtained from the sum of the converged value with
basis up to $h$-symmetry (see the convergence Table \ref{con-alpha}) and 
the estimated contribution from $i$, $j$ and $k$-symmetries. The combine
contribution from $i$, $j$ and $k$-symmetries are $\approx$ 0.5\%,
0.07\% and 0.02\%, respectively, for Cn, Nh$^+$ and Og. Though the 
contributions from the virtuals beyond $k$-symmetry is expected to be 
much lower, we select the highest contribution of 0.5\% from the case of 
Cn and attribute this as an upper bound to this source of uncertainty. 
The second source is the truncation of the dressed operator 
$\bar{\mathbf{D}}$ in Eq. (\ref{dbar}) to second order in $T^{(0)}$. 
In our previous work \cite{mani-10}, we concluded that the contribution 
from the remaining higher order terms is less than 0.1\%. So, from this 
source of uncertainty we consider 0.1\% as an upper bound. The third source 
is the partial inclusion of triple excitations in PRCC theory. The partial 
triples contribute $\approx$ 1.9\%, 0.04\%, 0.5\% of the PRCC value for Cn, 
Nh$^+$ and Og, respectively.  Since the perturbative triples account for 
the dominant contribution, we choose the highest contribution of $1.9$\% 
from Cn and take it as an upper bound. And, the last source of theoretical 
uncertainty is associated with the frequency-dependent Breit interaction 
which is not included in the present work. To estimate an upper bound to 
from source we use our previous work \cite{chattopadhyay-14} where using 
GRASP2K calculations we estimated an upper bound of $0.13$\% for Ra. Using 
this with the Breit contributions, we derive $\approx$ 0.62\%, 0.45\% and 
0.18\% as the contributions to Cn, Nh$^+$ and Og, respectively. Among these, 
we select the highest contribution of 0.62\% from the case of Cn and attribute 
this as an upper bound. There could be other sources of theoretical uncertainty, 
such as the higher order coupled perturbation of vacuum polarization and 
self-energy terms, quadruply and higher excited cluster operators, etc. These, 
however, have much lower contributions and their combined uncertainty could 
be below 0.1\%. Finally, combining the upper bounds of all four sources of 
uncertainties, we estimate a theoretical uncertainty of 3.2\% in the 
recommended values of $\alpha$.

\section{Conclusion}

  We have employed a fully relativistic coupled-cluster theory to compute the 
ground state electric dipole polarizability and electron correlation energy of 
SHEs Cn, Nh$^+$ and Og.  
In addition, to understand the trend of electron correlation as function of $Z$, 
we have calculated the correlation energies of three lighter homologs for each 
SHEs. To improve the accuracy of our results, contributions from the Breit 
interaction, QED corrections and partial triple excitations are also 
included. Moreover, in all calculations, very large bases up to $l$-symmetry
are used to check the convergence of the results.

Our recommended values of $\alpha$ for SHEs lie between the previous
results, more closer to the values from CCSD(T) \cite{pershina-08a,jerabek-18} 
and RPA \cite{dzuba-16} calculations. From our calculations we find that the
dominant contribution to $\alpha$ comes from the valence electrons, viz, 
$7s_{1/2}$ for both Cn and Nh$^+$, and $7p_{3/2}$ for Og. While $7s_{1/2}$ 
contributes more than 50\% of the total value for Cn and Nh$^+$, the 
contribution from $7p_{3/2}$ orbital to Og is more than 95\%. This could 
be attributed to the larger radial extent of these orbitals. 

From the analysis of electron correlation effects, we find that, for all
three groups, the core polarization contribution decreases as function of Z
for the lighter homologs. For SHEs, however, we observed an increased 
contribution, of the order of the first element considered in each group.
For the corrections from the Breit interaction, except for Cn and Og, a trend 
of decreasing contributions as function of Z is observed. On contrary, for 
Uehling potential and the self-energy corrections, we observed a trend of 
increasing contributions from lighter homologs to SHEs. The largest contributions 
from the Uehling potential are $\approx 0.2$\%, $0.3$\% and $0.3$\% of PRCC value 
for Cn, Nh$^+$ and Og, respectively. And, the same from the self-energy 
corrections are $\approx 0.4$\%, $0.2$\% and $0.1$\%, respectively.
The combined Breit+QED corrections to $\alpha$ are observed to be 
$\approx 0.5$\%, $0.4$\% and $0.6$\% for Cn, Nh$^+$ and Og, respectively.
Considering the importance of accurate properties results of SHEs, these are 
significant contributions, and can not be neglected.

We report the first result on the electron correlation energy for SHEs Cn, 
Nh$^+$ and Og using RCC. From our detailed analysis on correlation energy, we 
find that the second-order MBPT calculations overestimate the electron 
correlation energy for all superheavy elements and lighter homologs considered 
in this work.

\begin{acknowledgments}
We would like to thank Chandan Kumar Vishwakarma for useful discussions.
One of the authors, BKM, acknowledges the funding support from the
SERB (ECR/2016/001454). The results presented in the paper are based on the
computations using the High Performance Computing cluster, Padum, at the Indian
Institute of Technology Delhi, New Delhi.
\end{acknowledgments}

\newpage
\appendix

\section{Convergence table for $\alpha$}

In Table \ref{con-alpha} we provide the convergence trend of $\alpha$
as function of basis size. As it is evident from the table, the value
of $\alpha$ converges to $10^{-3}$ a.u. for all three SHEs.
\begin{table}[h!]
  \caption{Convergence trend of $\alpha$ calculated using the 
	Dirac-Coulomb Hamiltonian as a function of basis size.} 
  \label{con-alpha}
  \begin{ruledtabular}
  \begin{tabular}{ccc}
   Basis & Orbitals  &  $\alpha$ \\
       \hline
       \multicolumn{3}{c}{Cn}\\                 
       \hline  
  $90$  &  14s, 11p,  10d,  8f,  6g, 3h &    $32.772$   \\         
  $112$ &  16s, 13p,  12d, 10f,  8g, 5h &    $28.884$   \\       
  $132$ &  18s, 15p,  14d, 12f,  9g, 7h &    $28.094$   \\  
  $152$ &  20s, 17p,  16d, 14f, 11g, 8h &    $27.418$   \\         
  $172$ &  22s, 19p,  18d, 16f, 13g, 9h &    $27.116$   \\             
  $181$ &  23s, 20p,  19d, 17f, 14g, 9h &    $27.078$   \\         
  $191$ &  25s, 21p,  20d, 18f, 15g, 9h &    $26.948$   \\                   
  $200$ &  26s, 22p,  21d, 19f, 16g, 9h &    $26.944$   \\  
       \multicolumn{3}{c}{Nh$^+$}\\                 
       \hline
  $101$  & 15s, 13p, 12d,  8f,  5g,  5h  &    $19.493$   \\
  $123$  & 17s, 15p, 14d, 10f,  7g,  7h  &    $18.009$   \\        
  $145$  & 19s, 17p, 16d, 12f,  9g,  9h  &    $17.889$   \\ 
  $167$  & 21s, 19p, 18d, 14f, 11g, 11h  &    $17.330$   \\  
  $176$  & 22s, 20p, 19d, 15f, 12g, 11h  &    $17.155$   \\         
  $185$  & 23s, 21p, 20d, 16f, 13g, 11h  &    $17.082$   \\
  $194$  & 24s, 22p, 21d, 17f, 14g, 11h  &    $17.053$   \\                           
  $201$  & 25s, 23p, 22d, 18f, 14g, 11h  &    $17.053$   \\                           
       \multicolumn{3}{c}{Og}\\                 
       \hline 
  $86$   & 14s, 12p,  9d,  7f,  5g,  3h  &    $67.613$   \\
  $108$  & 16s, 14p, 11d,  9f,  7g,  5h  &    $65.556$   \\        
  $130$  & 18s, 16p, 13d, 11f,  9g,  7h  &    $60.134$   \\ 
  $152$  & 20s, 18p, 15d, 13f, 11g,  9h  &    $57.149$   \\  
  $170$  & 22s, 20p, 17d, 15f, 12g, 10h  &    $56.170$   \\         
  $179$  & 23s, 21p, 18d, 16f, 13g, 10h  &    $55.943$   \\
  $189$  & 25s, 22p, 19d, 17f, 14g, 10h  &    $55.941$   \\ 
 \end{tabular}  
 \end{ruledtabular}
\end{table}            
\section{Single-electron energies}

  The single-electron energies of GTOs for SHEs Cn, Nh$^+$ and Og are listed 
in the Tables \ref{ene-cn}, \ref{ene-nh} and \ref{ene-og}, respectively,
and compared with the numerical data from GRASP2k \cite{jonsson-13} and 
the energies from the B-spline \cite{} basis. In Table \ref{delta-en}, we
list the contributions from the Breit interaction, Uehling potential and 
the self-energy corrections to the single-electron energies.
\begin{table}[h]
        \caption{Orbital energies for core orbitals (in Hartree) from GTO 
         is compared with the GRASP2K and B-spline energies for Cn. 
         Here [x] represents multiplication by ${10^x}$.}
        \label{ene-cn}
        \begin{ruledtabular}
        \begin{tabular}{lrrr}
              {Orbital}
              & \multicolumn{1}{c}{\textrm{GRASP2K}}     &
                \multicolumn{1}{c}{\textrm{B-spline}}         &
                \multicolumn{1}{c}{\text{GTO}} \\     
            \hline
          $1s_{1/2}$ & 7070.83320  &  7071.11186  &  7070.83326    \\         
          $2s_{1/2}$ & 1444.87110  &  1444.92899  &  1444.87138    \\         
          $3s_{1/2}$ &  390.81374  &   390.82783  &   390.81406    \\         
          $4s_{1/2}$ &  113.44660  &   113.45069  &   113.44692    \\         
          $5s_{1/2}$ &   30.05645  &    30.05764  &    30.05670    \\         
          $6s_{1/2}$ &    5.68070  &     5.68099  &     5.68070    \\         
          $7s_{1/2}$ &    0.45115  &     0.45119  &     0.45114    \\ 
                                       \hline                                   
          $2p_{1/2}$ & 1405.71950  &  1405.72953   &  1405.71920    \\        
          $3p_{1/2}$ &  371.98098  &   371.98361   &   371.98109    \\        
          $4p_{1/2}$ &  104.23703  &   104.23780   &   104.23723    \\        
          $5p_{1/2}$ &   25.88719  &    25.88738   &    25.88726    \\        
          $6p_{1/2}$ &    4.12316  &     4.12320   &     4.12312    \\ 
                                       \hline                                   
          $2p_{3/2}$ & 1007.09780  &  1007.09651   &  1007.09804    \\        
          $3p_{3/2}$ &  274.99360  &   274.99302   &   274.99388    \\        
          $4p_{3/2}$ &   76.37753  &    76.37735   &    76.37776    \\        
          $5p_{3/2}$ &   17.99726  &    17.99719   &    17.99718    \\        
          $6p_{3/2}$ &    2.41564  &     2.41562   &     2.41564    \\ 
                                       \hline                                   
          $3d_{3/2}$ &  245.88774  &   245.88719   &   245.88802    \\        
          $4d_{3/2}$ &   62.08615  &    62.08599   &    62.08641    \\        
          $5d_{3/2}$ &   11.85266  &    11.85260   &    11.85259    \\         
          $6d_{3/2}$ &    0.56273  &     0.56271   &     0.56273    \\ 
                                       \hline                                   
          $3d_{5/2}$ &  229.40040  &   229.39991   &   229.40069    \\        
          $4d_{5/2}$ &   57.56171  &    57.56155   &    57.56196    \\        
          $5d_{5/2}$ &   10.70702  &    10.70697   &    10.70692    \\         
          $6d_{5/2}$ &    0.44208  &     0.44207   &     0.44208    \\ 
                                       \hline                                   
          $4f_{5/2}$ &   38.82989  &    38.82975   &    38.83023    \\        
          $5f_{5/2}$ &    3.33495  &     3.33492   &     3.33493    \\         
                                       \hline                                   
          $4f_{7/2}$ &   37.51594  &    37.51581   &    37.51628    \\        
          $5f_{7/2}$ &    3.09251  &     3.09247   &     3.09248    \\
        \end{tabular}
        \end{ruledtabular}
\end{table}            
\begin{table}[h]
        \caption{Orbital energies for core orbitals (in Hartree) from GTO 
         is compared with the GRASP2K and B-spline energies for Nh$^+$. 
         Here [x] represents multiplication by ${10^x}$.} 
        \label{ene-nh}
        \begin{ruledtabular}
        \begin{tabular}{lrrr}
              {Orbital}
              & \multicolumn{1}{c}{\textrm{GRASP2K}}     &
                \multicolumn{1}{c}{\textrm{B-spline}}    &
                \multicolumn{1}{c}{\text{GTO}} \\    
            \hline
          $1s_{1/2}$ &  7245.8727391  &  7246.182976  &  7245.873218  \\
          $2s_{1/2}$ &  1487.4479289  &  1487.512728  &  1487.448327  \\
          $3s_{1/2}$ &   403.5128636  &   403.529178  &   403.513164  \\
          $4s_{1/2}$ &   117.9615376  &   117.966271  &   117.961704  \\
          $5s_{1/2}$ &    31.8304473  &    31.831827  &    31.830498  \\
          $6s_{1/2}$ &     6.4543326  &     6.454659  &     6.454278  \\
          $7s_{1/2}$ &     0.8293919  &     0.829453  &     0.829389  \\ 
                                           \hline                    
          $2p_{1/2}$ &  1448.2662707  &  1448.277353  &  1448.266677  \\
          $3p_{1/2}$ &   384.4936370  &   384.497093  &   384.493911  \\
          $4p_{1/2}$ &   108.6076330  &   108.608617  &   108.607744  \\
          $5p_{1/2}$ &    27.5681904  &    27.568442  &    27.568231  \\
          $6p_{1/2}$ &     4.8384890  &     4.838538  &     4.838465  \\ 
                                           \hline                    
          $2p_{3/2}$ &  1028.6629340  &  1028.661537  &  1028.663316  \\
          $3p_{3/2}$ &   282.2280516  &   282.227605  &   282.228310  \\
          $4p_{3/2}$ &    79.1393402  &    79.139197  &    79.139459  \\
          $5p_{3/2}$ &    19.1605082  &    19.160449  &    19.160489  \\
          $6p_{3/2}$ &     2.9774011  &     2.977385  &     2.977406  \\ 
                                           \hline                    
          $3d_{3/2}$ &   252.7000748  &   252.699678  &   252.700378  \\
          $4d_{3/2}$ &    64.6075721  &    64.607451  &    64.607745  \\
          $5d_{3/2}$ &    12.8763169  &    12.876269  &    12.876364  \\ 
          $6d_{3/2}$ &     1.0204591  &     1.020450  &     1.020479  \\ 
                                           \hline                    
          $3d_{5/2}$ &   235.5220171  &   235.521671  &   235.522311  \\
          $4d_{5/2}$ &    59.8711971  &    59.871089  &    59.871368  \\
          $5d_{5/2}$ &    11.6644138  &    11.664335  &    11.664411  \\ 
          $6d_{5/2}$ &     0.8811957  &     0.881171  &     0.881195  \\ 
                                           \hline                    
          $4f_{5/2}$ &    40.8372335  &    40.837027  &    40.837249  \\
          $5f_{5/2}$ &     4.0945793  &     4.094568  &     4.094589  \\ 
                                           \hline                    
          $4f_{7/2}$ &    39.4570885  &    39.456886  &    39.457103  \\
          $5f_{7/2}$ &     3.8335434  &     3.833539  &     3.833556  \\ 
        \end{tabular}  
        \end{ruledtabular}
\end{table}                          
\begin{table}[h]
        \caption{Orbital energies for core orbitals (in Hartree) from GTO 
         is compared with the GRASP2K and B-spline energies for Og. 
         Here [x] represents multiplication by ${10^x}$.} 
        \label{ene-og}
        \begin{ruledtabular}
        \begin{tabular}{lrrr}
              {Orbital}
              & \multicolumn{1}{c}{\textrm{GRASP2K}}     &
                \multicolumn{1}{c}{\textrm{B-spline}}         &
                 \multicolumn{1}{c}{\text{GTO}} \\    
            \hline
          $1s_{1/2}$ & 8185.36230  &  8185.93230  &  8185.36258    \\
          $2s_{1/2}$ & 1718.80780  &  1718.93698  &  1718.80803    \\
          $3s_{1/2}$ &  471.19401  &   471.22553  &   471.19411    \\
          $4s_{1/2}$ &  140.97641  &   140.98548  &   140.97632    \\
          $5s_{1/2}$ &   39.88519  &   39.88767   &    39.88495    \\
          $6s_{1/2}$ &    8.98686  &     8.98760  &     8.98678    \\
          $7s_{1/2}$ &    1.29699  &     1.29711  &     1.29696    \\ 
                                  \hline 
          $2p_{1/2}$ & 1681.71710  &  1681.74523  &  1681.71618    \\
          $3p_{1/2}$ &  451.72699  &   451.73439  &   451.72665    \\
          $4p_{1/2}$ &  131.02105  &   131.02303  &   131.02069    \\
          $5p_{1/2}$ &   35.18375  &    35.18404  &    35.18340    \\
          $6p_{1/2}$ &    7.07694  &     7.07713  &     7.07689    \\
          $7p_{1/2}$ &    0.73956  &     0.73948  &     0.73944    \\ 
                                  \hline
          $2p_{3/2}$ &  113.85447  &  1138.54073  &  1138.54500   \\
          $3p_{3/2}$ &  318.33517  &   318.33345  &   318.33518    \\
          $4p_{3/2}$ &   92.02425  &    92.02349  &    92.02406    \\
          $5p_{3/2}$ &   23.66280  &    23.66234  &    23.66242    \\
          $6p_{3/2}$ &    4.21643  &     4.21630  &     4.21633    \\ 
          $7p_{3/2}$ &    0.30564  &     0.30564  &     0.30565    \\ 
                                  \hline 
          $3d_{3/2}$ &  286.65027  &   286.64895  &   286.65036    \\
          $4d_{3/2}$ &   76.26542  &    76.26467  &    76.26535    \\
          $5d_{3/2}$ &   16.66319  &    16.66277  &    16.66297    \\ 
          $6d_{3/2}$ &    1.76398  &     1.76387  &     1.76394    \\ 
                                  \hline 
          $3d_{5/2}$ &  265.67617  &   265.67496  &   265.67625    \\
          $4d_{5/2}$ &   70.35026  &    70.34956  &    70.35019    \\
          $5d_{5/2}$ &   15.07066  &    15.07028  &    15.07044    \\ 
          $6d_{5/2}$ &    1.49296  &     1.49285  &     1.49291    \\ 
                                  \hline 
          $4f_{5/2}$ &   49.79167  &    49.79139  &    49.79205    \\
          $5f_{5/2}$ &    6.51102  &     6.51097  &     6.51114    \\ 
                                  \hline 
          $4f_{7/2}$ &   48.04247  &    48.04220  &    48.04286    \\
          $5f_{7/2}$ &    6.14074  &     6.14070  &     6.14086    \\ 
        \end{tabular}  
        \end{ruledtabular}
\end{table}                          
\begin{table*}[h]
     \caption{The orbital energies for core orbitals from vacuum polarization and
     self energy correction for Cn, Nh$^+$ and Og.}
     \label{delta-en}
     \begin{ruledtabular}
     \begin{tabular}{lcccccccccccc}
       & &   Cn  &&&& Nh$^+$  &&&& Og \\
       \cline{2-5}  \cline{6-9} \cline{10-13}
  Orbital & \multicolumn{2}{c}{$\Delta\epsilon_{\rm Ue}$} & \multicolumn{2}{c}{$\Delta\epsilon_{\rm SE}$}
  & \multicolumn{2}{c}{$\Delta\epsilon_{\rm Ue}$} & \multicolumn{2}{c}{$\Delta\epsilon_{\rm SE}$}
  &\multicolumn{2}{c}{$\Delta\epsilon_{\rm Ue}$} & \multicolumn{2}{c}{$\Delta\epsilon_{\rm SE}$}  \\
  \cline{2-3} \cline{4-5}     \cline{6-7} \cline{8-9}     \cline{10-11} \cline{12-13}
   & \multicolumn{1}{c}{\textrm{Ours}} & \multicolumn{1}{c}{\textrm{Ref. \cite{karol-18b}}}
   & \multicolumn{1}{c}{\textrm{Ours}} & \multicolumn{1}{c}{\textrm{Ref. \cite{karol-18b}}}
   & \multicolumn{1}{c}{\textrm{Ours}} & \multicolumn{1}{c}{\textrm{}}
   & \multicolumn{1}{c}{\textrm{Ours}} & \multicolumn{1}{c}{\textrm{}}
   & \multicolumn{1}{c}{\textrm{Ours}} & \multicolumn{1}{c}{\textrm{Ref. \cite{karol-18b}}}
   & \multicolumn{1}{c}{\textrm{Ours}} & \multicolumn{1}{c}{\textrm{Ref. \cite{karol-18b}}}      \\
         \hline
   $1s_{1/2}$&-11.1193&-11.4416&30.6243&30.5752&-11.8402&&31.9902&&-16.2622&-16.7082&39.9045&39.7825  \\
   $2s_{1/2}$& -2.1505&- 2.2283& 5.9196& 5.7672& -2.3160&& 6.2339&& -3.3693& -3.4810& 8.1051& 7.7743  \\
   $3s_{1/2}$& -0.5193&- 0.5410& 1.4424& 1.5061& -0.5597&& 1.5198&& -0.8134& -0.8487& 1.9781& 2.1609  \\
   $4s_{1/2}$& -0.1494&- 0.1560& 0.4165& 0.4375& -0.1614&& 0.4400&& -0.2377& -0.2482& 0.5795& 0.6369  \\
   $5s_{1/2}$& -0.0436&        & 0.1214&       & -0.0474&& 0.1291&& -0.0718&        & 0.1783&         \\
   $6s_{1/2}$& -0.0108&        & 0.0297&       & -0.0119&& 0.0322&& -0.0197&        & 0.0500&         \\
   $7s_{1/2}$& -0.0015&        & 0.0037&       & -0.0019&& 0.0047&& -0.0040&        & 0.0098&         \\
                                \hline
   $2p_{1/2}$&-0.5505&-0.6632&1.6011&1.6611&-0.6457&&1.7455&&-1.0858&-1.2661&2.7124&2.7853  \\
   $3p_{1/2}$&-0.1486&-0.1801&0.4445&0.5091&-0.1662&&0.4835&&-0.2926&-0.3430&0.7399&0.8812  \\
   $4p_{1/2}$&-0.0422&-0.0523&0.1278&0.1551&-0.0473&&0.1394&&-0.0845&-0.1010&0.2156&0.2712  \\
   $5p_{1/2}$&-0.0115&       &0.0359&      &-0.0130&&0.0395&&-0.0242&       &0.0674&        \\
   $6p_{1/2}$&-0.0024&       &0.0080&      &-0.0028&&0.0090&&-0.0059&       &0.0177&        \\
   $7p_{1/2}$&       &       &      &      &       &&      &&-0.0010&       &0.0027&        \\
                                \hline
   $2p_{3/2}$&0.0554&-0.0122&0.6914&0.6997&0.0594&&0.7207&&0.0848&-0.0169&0.8799&0.9013  \\
   $3p_{3/2}$&0.0183&-0.0038&0.1896&0.2002&0.0197&&0.1983&&0.0287&-0.0054&0.2476&0.2644  \\
   $4p_{3/2}$&0.0068&-0.0011&0.0565&0.0613&0.0074&&0.0594&&0.0110&-0.0017&0.0774&0.0827  \\
   $5p_{3/2}$&0.0026&       &0.0157&      &0.0028&&0.0167&&0.0043&       &0.0224&        \\
   $6p_{3/2}$&0.0009&       &0.0029&      &0.0010&&0.0032&&0.0016&       &0.0046&        \\
   $7p_{3/2}$&      &       &      &      &      &&      &&      &0.0003 &0.0005& 0.0445       \\
                                \hline
   $3d_{3/2}$&0.0205&-0.0001&-0.0015&-0.0014&0.0221&&-0.0012    &  &0.0316&-0.0002&0.0012&0.0012  \\
   $4d_{3/2}$&0.0070&   0   &-0.0005& 0.0009&0.0076&&-0.0004    &  &0.0112&-0.0001&0.0004&0.0023  \\
   $5d_{3/2}$&0.0024&       &-0.0001&       &0.0026&&-0.0001    &  &0.0040&       &0.0001&        \\
   $6d_{3/2}$&0.0007&       & 0     &       &0.0007&&    0  & &0.0011 &          & 0      &          \\
                                \hline
   $3d_{5/2}$&0.0185&  0    &0.0253&0.0231&0.0199&&0.0265&&0.0282&  0   &0.0333&0.0304  \\
   $4d_{5/2}$&0.0064&  0    &0.0084&0.0064&0.0069&&0.0089&&0.0102&  0   &0.0115&0.0086  \\
   $5d_{5/2}$&0.0022&       &0.0024&      &0.0024&&0.0023&&0.0036&      &0.0031&        \\
   $6d_{5/2}$&0.0006&       &0.0002&      &0.0006&&0.0003&&0.0010&      &0.0005&        \\
                                \hline
   $4f_{5/2}$&0.0055&&& &0.0059&  & &  &  0.0088 &         &         &          \\
   $5f_{5/2}$&0.0015&&& &0.0017&  & &  &  0.0027 &         &         &          \\
                                \hline
   $4f_{7/2}$&0.0053& & & & 0.0057 &  & &  & 0.0084 & & &          \\
   $5f_{7/2}$&0.0015& & & & 0.0016 &  & &  & 0.0026 & & &          \\
\end{tabular}
\end{ruledtabular}
\end{table*}
 

\bibliography{super_heavy}

\begin{thebibliography}{46}%
\makeatletter
\providecommand \@ifxundefined [1]{%
 \@ifx{#1\undefined}
}%
\providecommand \@ifnum [1]{%
 \ifnum #1\expandafter \@firstoftwo
 \else \expandafter \@secondoftwo
 \fi
}%
\providecommand \@ifx [1]{%
 \ifx #1\expandafter \@firstoftwo
 \else \expandafter \@secondoftwo
 \fi
}%
\providecommand \natexlab [1]{#1}%
\providecommand \enquote  [1]{``#1''}%
\providecommand \bibnamefont  [1]{#1}%
\providecommand \bibfnamefont [1]{#1}%
\providecommand \citenamefont [1]{#1}%
\providecommand \href@noop [0]{\@secondoftwo}%
\providecommand \href [0]{\begingroup \@sanitize@url \@href}%
\providecommand \@href[1]{\@@startlink{#1}\@@href}%
\providecommand \@@href[1]{\endgroup#1\@@endlink}%
\providecommand \@sanitize@url [0]{\catcode `\\12\catcode `\$12\catcode
  `\&12\catcode `\#12\catcode `\^12\catcode `\_12\catcode `\%12\relax}%
\providecommand \@@startlink[1]{}%
\providecommand \@@endlink[0]{}%
\providecommand \url  [0]{\begingroup\@sanitize@url \@url }%
\providecommand \@url [1]{\endgroup\@href {#1}{\urlprefix }}%
\providecommand \urlprefix  [0]{URL }%
\providecommand \Eprint [0]{\href }%
\providecommand \doibase [0]{http://dx.doi.org/}%
\providecommand \selectlanguage [0]{\@gobble}%
\providecommand \bibinfo  [0]{\@secondoftwo}%
\providecommand \bibfield  [0]{\@secondoftwo}%
\providecommand \translation [1]{[#1]}%
\providecommand \BibitemOpen [0]{}%
\providecommand \bibitemStop [0]{}%
\providecommand \bibitemNoStop [0]{.\EOS\space}%
\providecommand \EOS [0]{\spacefactor3000\relax}%
\providecommand \BibitemShut  [1]{\csname bibitem#1\endcsname}%
\let\auto@bib@innerbib\@empty
\bibitem [{\citenamefont {Türler}\ and\ \citenamefont
  {Pershina}(2013)}]{turler-13}%
  \BibitemOpen
  \bibfield  {author} {\bibinfo {author} {\bibfnamefont {Andreas}\ \bibnamefont
  {Türler}}\ and\ \bibinfo {author} {\bibfnamefont {Valeria}\ \bibnamefont
  {Pershina}},\ }\bibfield  {title} {\enquote {\bibinfo {title} {Advances in
  the production and chemistry of the heaviest elements},}\ }\href {\doibase
  10.1021/cr3002438} {\bibfield  {journal} {\bibinfo  {journal} {Chemical
  Reviews}\ }\textbf {\bibinfo {volume} {113}},\ \bibinfo {pages} {1237--1312}
  (\bibinfo {year} {2013})},\ \bibinfo {note} {pMID: 23402305},\ \Eprint
  {http://arxiv.org/abs/https://doi.org/10.1021/cr3002438}
  {https://doi.org/10.1021/cr3002438} \BibitemShut {NoStop}%
\bibitem [{\citenamefont {Schädel}(2015)}]{matthias-15}%
  \BibitemOpen
  \bibfield  {author} {\bibinfo {author} {\bibfnamefont {Matthias}\
  \bibnamefont {Schädel}},\ }\bibfield  {title} {\enquote {\bibinfo {title}
  {Chemistry of the superheavy elements},}\ }\href {\doibase
  10.1098/rsta.2014.0191} {\bibfield  {journal} {\bibinfo  {journal}
  {Philosophical Transactions of the Royal Society A: Mathematical, Physical
  and Engineering Sciences}\ }\textbf {\bibinfo {volume} {373}},\ \bibinfo
  {pages} {20140191} (\bibinfo {year} {2015})},\ \Eprint
  {http://arxiv.org/abs/https://royalsocietypublishing.org/doi/pdf/10.1098/rsta.2014.0191}
  {https://royalsocietypublishing.org/doi/pdf/10.1098/rsta.2014.0191}
  \BibitemShut {NoStop}%
\bibitem [{\citenamefont {Pershina}(2015)}]{pershina-15}%
  \BibitemOpen
  \bibfield  {author} {\bibinfo {author} {\bibfnamefont {V.}~\bibnamefont
  {Pershina}},\ }\bibfield  {title} {\enquote {\bibinfo {title} {Electronic
  structure and properties of superheavy elements},}\ }\href {\doibase
  https://doi.org/10.1016/j.nuclphysa.2015.04.007} {\bibfield  {journal}
  {\bibinfo  {journal} {Nuclear Physics A}\ }\textbf {\bibinfo {volume}
  {944}},\ \bibinfo {pages} {578 -- 613} (\bibinfo {year} {2015})},\ \bibinfo
  {note} {special Issue on Superheavy Elements}\BibitemShut {NoStop}%
\bibitem [{\citenamefont {Schwerdtfeger}\ \emph {et~al.}(2015)\citenamefont
  {Schwerdtfeger}, \citenamefont {Pašteka}, \citenamefont {Punnett},\ and\
  \citenamefont {Bowman}}]{peter-15}%
  \BibitemOpen
  \bibfield  {author} {\bibinfo {author} {\bibfnamefont {Peter}\ \bibnamefont
  {Schwerdtfeger}}, \bibinfo {author} {\bibfnamefont {Lukáš~F.}\ \bibnamefont
  {Pašteka}}, \bibinfo {author} {\bibfnamefont {Andrew}\ \bibnamefont
  {Punnett}}, \ and\ \bibinfo {author} {\bibfnamefont {Patrick~O.}\
  \bibnamefont {Bowman}},\ }\bibfield  {title} {\enquote {\bibinfo {title}
  {Relativistic and quantum electrodynamic effects in superheavy elements},}\
  }\href {\doibase https://doi.org/10.1016/j.nuclphysa.2015.02.005} {\bibfield
  {journal} {\bibinfo  {journal} {Nuclear Physics A}\ }\textbf {\bibinfo
  {volume} {944}},\ \bibinfo {pages} {551 -- 577} (\bibinfo {year} {2015})},\
  \bibinfo {note} {special Issue on Superheavy Elements}\BibitemShut {NoStop}%
\bibitem [{\citenamefont {Eliav}\ \emph {et~al.}(2015)\citenamefont {Eliav},
  \citenamefont {Fritzsche},\ and\ \citenamefont {Kaldor}}]{eliav-15}%
  \BibitemOpen
  \bibfield  {author} {\bibinfo {author} {\bibfnamefont {Ephraim}\ \bibnamefont
  {Eliav}}, \bibinfo {author} {\bibfnamefont {Stephan}\ \bibnamefont
  {Fritzsche}}, \ and\ \bibinfo {author} {\bibfnamefont {Uzi}\ \bibnamefont
  {Kaldor}},\ }\bibfield  {title} {\enquote {\bibinfo {title} {Electronic
  structure theory of the superheavy elements},}\ }\href {\doibase
  https://doi.org/10.1016/j.nuclphysa.2015.06.017} {\bibfield  {journal}
  {\bibinfo  {journal} {Nuclear Physics A}\ }\textbf {\bibinfo {volume}
  {944}},\ \bibinfo {pages} {518 -- 550} (\bibinfo {year} {2015})},\ \bibinfo
  {note} {special Issue on Superheavy Elements}\BibitemShut {NoStop}%
\bibitem [{\citenamefont {Giuliani}\ \emph {et~al.}(2019)\citenamefont
  {Giuliani}, \citenamefont {Matheson}, \citenamefont {Nazarewicz},
  \citenamefont {Olsen}, \citenamefont {Reinhard}, \citenamefont {Sadhukhan},
  \citenamefont {Schuetrumpf}, \citenamefont {Schunck},\ and\ \citenamefont
  {Schwerdtfeger}}]{giuliani-19}%
  \BibitemOpen
  \bibfield  {author} {\bibinfo {author} {\bibfnamefont {S.~A.}\ \bibnamefont
  {Giuliani}}, \bibinfo {author} {\bibfnamefont {Z.}~\bibnamefont {Matheson}},
  \bibinfo {author} {\bibfnamefont {W.}~\bibnamefont {Nazarewicz}}, \bibinfo
  {author} {\bibfnamefont {E.}~\bibnamefont {Olsen}}, \bibinfo {author}
  {\bibfnamefont {P.-G.}\ \bibnamefont {Reinhard}}, \bibinfo {author}
  {\bibfnamefont {J.}~\bibnamefont {Sadhukhan}}, \bibinfo {author}
  {\bibfnamefont {B.}~\bibnamefont {Schuetrumpf}}, \bibinfo {author}
  {\bibfnamefont {N.}~\bibnamefont {Schunck}}, \ and\ \bibinfo {author}
  {\bibfnamefont {P.}~\bibnamefont {Schwerdtfeger}},\ }\bibfield  {title}
  {\enquote {\bibinfo {title} {Colloquium: Superheavy elements: Oganesson and
  beyond},}\ }\href {\doibase 10.1103/RevModPhys.91.011001} {\bibfield
  {journal} {\bibinfo  {journal} {Rev. Mod. Phys.}\ }\textbf {\bibinfo {volume}
  {91}},\ \bibinfo {pages} {011001} (\bibinfo {year} {2019})}\BibitemShut
  {NoStop}%
\bibitem [{\citenamefont {Shaughnessy}\ and\ \citenamefont
  {Schadel}(2014)}]{schadel-14}%
  \BibitemOpen
  \bibfield  {author} {\bibinfo {author} {\bibfnamefont {D.}~\bibnamefont
  {Shaughnessy}}\ and\ \bibinfo {author} {\bibfnamefont {M.}~\bibnamefont
  {Schadel}},\ }\href@noop {} {\emph {\bibinfo {title} {The Chemistry of the
  Superheavy Elements}}},\ \bibinfo {edition} {2nd}\ ed.\ (\bibinfo
  {publisher} {Springer},\ \bibinfo {address} {Heidelberg},\ \bibinfo {year}
  {2014})\BibitemShut {NoStop}%
\bibitem [{\citenamefont {Pershina}\ and\ \citenamefont
  {Hoffman}(2008)}]{hoffman-06}%
  \BibitemOpen
  \bibfield  {author} {\bibinfo {author} {\bibfnamefont {V.}~\bibnamefont
  {Pershina}}\ and\ \bibinfo {author} {\bibfnamefont {D.C.}\ \bibnamefont
  {Hoffman}},\ }\href@noop {} {\emph {\bibinfo {title} {Transactinide Elements
  and Future Elements}}}\ (\bibinfo  {publisher} {Springer, Dordrecht},\
  \bibinfo {year} {2008})\BibitemShut {NoStop}%
\bibitem [{\citenamefont {Khriplovich}(1991)}]{khriplovich-91}%
  \BibitemOpen
  \bibfield  {author} {\bibinfo {author} {\bibfnamefont {I.B.}\ \bibnamefont
  {Khriplovich}},\ }\href@noop {} {\emph {\bibinfo {title} {Parity
  Nonconservation in Atomic Phenomena}}}\ (\bibinfo  {publisher} {Gordon and
  Breach Science Publishers},\ \bibinfo {address} {Philadelphia},\ \bibinfo
  {year} {1991})\BibitemShut {NoStop}%
\bibitem [{\citenamefont {Griffith}\ \emph {et~al.}(2009)\citenamefont
  {Griffith}, \citenamefont {Swallows}, \citenamefont {Loftus}, \citenamefont
  {Romalis}, \citenamefont {Heckel},\ and\ \citenamefont
  {Fortson}}]{griffith-09}%
  \BibitemOpen
  \bibfield  {author} {\bibinfo {author} {\bibfnamefont {W.~C.}\ \bibnamefont
  {Griffith}}, \bibinfo {author} {\bibfnamefont {M.~D.}\ \bibnamefont
  {Swallows}}, \bibinfo {author} {\bibfnamefont {T.~H.}\ \bibnamefont
  {Loftus}}, \bibinfo {author} {\bibfnamefont {M.~V.}\ \bibnamefont {Romalis}},
  \bibinfo {author} {\bibfnamefont {B.~R.}\ \bibnamefont {Heckel}}, \ and\
  \bibinfo {author} {\bibfnamefont {E.~N.}\ \bibnamefont {Fortson}},\
  }\bibfield  {title} {\enquote {\bibinfo {title} {Improved limit on the
  permanent electric dipole moment of \mbox{Hg$^{199}$}},}\ }\href {\doibase
  10.1103/PhysRevLett.102.101601} {\bibfield  {journal} {\bibinfo  {journal}
  {Phys. Rev. Lett.}\ }\textbf {\bibinfo {volume} {102}},\ \bibinfo {pages}
  {101601} (\bibinfo {year} {2009})}\BibitemShut {NoStop}%
\bibitem [{\citenamefont {Udem}\ \emph {et~al.}(2002)\citenamefont {Udem},
  \citenamefont {Holzwarth},\ and\ \citenamefont {Hansch}}]{udem-02}%
  \BibitemOpen
  \bibfield  {author} {\bibinfo {author} {\bibfnamefont {Th.}\ \bibnamefont
  {Udem}}, \bibinfo {author} {\bibfnamefont {R.}~\bibnamefont {Holzwarth}}, \
  and\ \bibinfo {author} {\bibfnamefont {T.~W.}\ \bibnamefont {Hansch}},\
  }\bibfield  {title} {\enquote {\bibinfo {title} {Optical frequency
  metrology},}\ }\href {\doibase http://dx.doi.org/10.1038/416233a} {\bibfield
  {journal} {\bibinfo  {journal} {Nature}\ }\textbf {\bibinfo {volume} {416}},\
  \bibinfo {pages} {233} (\bibinfo {year} {2002})}\BibitemShut {NoStop}%
\bibitem [{\citenamefont {Lewenstein}\ \emph {et~al.}(1994)\citenamefont
  {Lewenstein}, \citenamefont {Balcou}, \citenamefont {Ivanov}, \citenamefont
  {L'Huillier},\ and\ \citenamefont {Corkum}}]{lewenstein-94}%
  \BibitemOpen
  \bibfield  {author} {\bibinfo {author} {\bibfnamefont {M.}~\bibnamefont
  {Lewenstein}}, \bibinfo {author} {\bibfnamefont {Ph.}\ \bibnamefont
  {Balcou}}, \bibinfo {author} {\bibfnamefont {M.~Yu.}\ \bibnamefont {Ivanov}},
  \bibinfo {author} {\bibfnamefont {Anne}\ \bibnamefont {L'Huillier}}, \ and\
  \bibinfo {author} {\bibfnamefont {P.~B.}\ \bibnamefont {Corkum}},\ }\bibfield
   {title} {\enquote {\bibinfo {title} {Theory of high-harmonic generation by
  low-frequency laser fields},}\ }\href {\doibase 10.1103/PhysRevA.49.2117}
  {\bibfield  {journal} {\bibinfo  {journal} {Phys. Rev. A}\ }\textbf {\bibinfo
  {volume} {49}},\ \bibinfo {pages} {2117--2132} (\bibinfo {year}
  {1994})}\BibitemShut {NoStop}%
\bibitem [{\citenamefont {Anderson}\ \emph {et~al.}(1995)\citenamefont
  {Anderson}, \citenamefont {Ensher}, \citenamefont {Matthews}, \citenamefont
  {Wieman},\ and\ \citenamefont {Cornell}}]{anderson-95}%
  \BibitemOpen
  \bibfield  {author} {\bibinfo {author} {\bibfnamefont {M.~H.}\ \bibnamefont
  {Anderson}}, \bibinfo {author} {\bibfnamefont {J.~R.}\ \bibnamefont
  {Ensher}}, \bibinfo {author} {\bibfnamefont {M.~R.}\ \bibnamefont
  {Matthews}}, \bibinfo {author} {\bibfnamefont {C.~E.}\ \bibnamefont
  {Wieman}}, \ and\ \bibinfo {author} {\bibfnamefont {E.~A.}\ \bibnamefont
  {Cornell}},\ }\bibfield  {title} {\enquote {\bibinfo {title} {Observation of
  \mbox{B}ose-\mbox{E}instein condensation in a dilute atomic vapor},}\ }\href
  {\doibase 10.1126/science.269.5221.198} {\bibfield  {journal} {\bibinfo
  {journal} {Science}\ }\textbf {\bibinfo {volume} {269}},\ \bibinfo {pages}
  {198--201} (\bibinfo {year} {1995})}\BibitemShut {NoStop}%
\bibitem [{\citenamefont {Karshenboim}\ and\ \citenamefont
  {Peik}(2010)}]{karshenboim-10}%
  \BibitemOpen
  \bibfield  {author} {\bibinfo {author} {\bibfnamefont {S.~G.}\ \bibnamefont
  {Karshenboim}}\ and\ \bibinfo {author} {\bibfnamefont {E}~\bibnamefont
  {Peik}},\ }\href@noop {} {\emph {\bibinfo {title} {Astrophysics, Clocks and
  Fundamental Constants, Lecture Notes in Physics}}}\ (\bibinfo  {publisher}
  {Springer},\ \bibinfo {address} {New York},\ \bibinfo {year}
  {2010})\BibitemShut {NoStop}%
\bibitem [{\citenamefont {Seth}\ \emph {et~al.}(1997)\citenamefont {Seth},
  \citenamefont {Schwerdtfeger},\ and\ \citenamefont {Dolg}}]{seth-97}%
  \BibitemOpen
  \bibfield  {author} {\bibinfo {author} {\bibfnamefont {Michael}\ \bibnamefont
  {Seth}}, \bibinfo {author} {\bibfnamefont {Peter}\ \bibnamefont
  {Schwerdtfeger}}, \ and\ \bibinfo {author} {\bibfnamefont {Michael}\
  \bibnamefont {Dolg}},\ }\bibfield  {title} {\enquote {\bibinfo {title} {The
  chemistry of the superheavy elements. i. pseudopotentials for 111 and 112 and
  relativistic coupled cluster calculations for (112)h+, (112)f2, and
  (112)f4},}\ }\href {\doibase 10.1063/1.473437} {\bibfield  {journal}
  {\bibinfo  {journal} {The Journal of Chemical Physics}\ }\textbf {\bibinfo
  {volume} {106}},\ \bibinfo {pages} {3623--3632} (\bibinfo {year} {1997})},\
  \Eprint {http://arxiv.org/abs/https://doi.org/10.1063/1.473437}
  {https://doi.org/10.1063/1.473437} \BibitemShut {NoStop}%
\bibitem [{\citenamefont {Nash}(2005)}]{nash-05}%
  \BibitemOpen
  \bibfield  {author} {\bibinfo {author} {\bibfnamefont {Clinton~S.}\
  \bibnamefont {Nash}},\ }\bibfield  {title} {\enquote {\bibinfo {title}
  {Atomic and molecular properties of elements 112, 114, and 118},}\ }\href
  {\doibase 10.1021/jp050736o} {\bibfield  {journal} {\bibinfo  {journal} {The
  Journal of Physical Chemistry A}\ }\textbf {\bibinfo {volume} {109}},\
  \bibinfo {pages} {3493--3500} (\bibinfo {year} {2005})},\ \bibinfo {note}
  {pMID: 16833687},\ \Eprint
  {http://arxiv.org/abs/https://doi.org/10.1021/jp050736o}
  {https://doi.org/10.1021/jp050736o} \BibitemShut {NoStop}%
\bibitem [{\citenamefont {Dzuba}(2016)}]{dzuba-16}%
  \BibitemOpen
  \bibfield  {author} {\bibinfo {author} {\bibfnamefont {V.~A.}\ \bibnamefont
  {Dzuba}},\ }\bibfield  {title} {\enquote {\bibinfo {title} {{Ionization
  potentials and polarizabilities of superheavy elements from Db to Cn
  (Z=105--112)}},}\ }\href {\doibase 10.1103/PhysRevA.93.032519} {\bibfield
  {journal} {\bibinfo  {journal} {Phys. Rev. A}\ }\textbf {\bibinfo {volume}
  {93}},\ \bibinfo {pages} {032519} (\bibinfo {year} {2016})}\BibitemShut
  {NoStop}%
\bibitem [{\citenamefont {Pershina}\ \emph
  {et~al.}(2008{\natexlab{a}})\citenamefont {Pershina}, \citenamefont
  {Borschevsky}, \citenamefont {Eliav},\ and\ \citenamefont
  {Kaldor}}]{pershina-08a}%
  \BibitemOpen
  \bibfield  {author} {\bibinfo {author} {\bibfnamefont {V.}~\bibnamefont
  {Pershina}}, \bibinfo {author} {\bibfnamefont {A.}~\bibnamefont
  {Borschevsky}}, \bibinfo {author} {\bibfnamefont {E.}~\bibnamefont {Eliav}},
  \ and\ \bibinfo {author} {\bibfnamefont {U.}~\bibnamefont {Kaldor}},\
  }\bibfield  {title} {\enquote {\bibinfo {title} {Prediction of the adsorption
  behavior of elements 112 and 114 on inert surfaces from ab initio
  dirac-coulomb atomic calculations},}\ }\href {\doibase 10.1063/1.2814242}
  {\bibfield  {journal} {\bibinfo  {journal} {The Journal of Chemical Physics}\
  }\textbf {\bibinfo {volume} {128}},\ \bibinfo {pages} {024707} (\bibinfo
  {year} {2008}{\natexlab{a}})},\ \Eprint
  {http://arxiv.org/abs/https://doi.org/10.1063/1.2814242}
  {https://doi.org/10.1063/1.2814242} \BibitemShut {NoStop}%
\bibitem [{\citenamefont {Pershina}\ \emph
  {et~al.}(2008{\natexlab{b}})\citenamefont {Pershina}, \citenamefont
  {Borschevsky}, \citenamefont {Eliav},\ and\ \citenamefont
  {Kaldor}}]{pershina-08b}%
  \BibitemOpen
  \bibfield  {author} {\bibinfo {author} {\bibfnamefont {V.}~\bibnamefont
  {Pershina}}, \bibinfo {author} {\bibfnamefont {A.}~\bibnamefont
  {Borschevsky}}, \bibinfo {author} {\bibfnamefont {E.}~\bibnamefont {Eliav}},
  \ and\ \bibinfo {author} {\bibfnamefont {U.}~\bibnamefont {Kaldor}},\
  }\bibfield  {title} {\enquote {\bibinfo {title} {Adsorption of inert gases
  including element 118 on noble metal and inert surfaces from ab initio
  dirac–coulomb atomic calculations},}\ }\href {\doibase 10.1063/1.2988318}
  {\bibfield  {journal} {\bibinfo  {journal} {The Journal of Chemical Physics}\
  }\textbf {\bibinfo {volume} {129}},\ \bibinfo {pages} {144106} (\bibinfo
  {year} {2008}{\natexlab{b}})},\ \Eprint
  {http://arxiv.org/abs/https://doi.org/10.1063/1.2988318}
  {https://doi.org/10.1063/1.2988318} \BibitemShut {NoStop}%
\bibitem [{\citenamefont {Jerabek}\ \emph {et~al.}(2018)\citenamefont
  {Jerabek}, \citenamefont {Schuetrumpf}, \citenamefont {Schwerdtfeger},\ and\
  \citenamefont {Nazarewicz}}]{jerabek-18}%
  \BibitemOpen
  \bibfield  {author} {\bibinfo {author} {\bibfnamefont {Paul}\ \bibnamefont
  {Jerabek}}, \bibinfo {author} {\bibfnamefont {Bastian}\ \bibnamefont
  {Schuetrumpf}}, \bibinfo {author} {\bibfnamefont {Peter}\ \bibnamefont
  {Schwerdtfeger}}, \ and\ \bibinfo {author} {\bibfnamefont {Witold}\
  \bibnamefont {Nazarewicz}},\ }\bibfield  {title} {\enquote {\bibinfo {title}
  {Electron and nucleon localization functions of oganesson: Approaching the
  thomas-fermi limit},}\ }\href {\doibase 10.1103/PhysRevLett.120.053001}
  {\bibfield  {journal} {\bibinfo  {journal} {Phys. Rev. Lett.}\ }\textbf
  {\bibinfo {volume} {120}},\ \bibinfo {pages} {053001} (\bibinfo {year}
  {2018})}\BibitemShut {NoStop}%
\bibitem [{\citenamefont {Chattopadhyay}\ \emph
  {et~al.}(2012{\natexlab{a}})\citenamefont {Chattopadhyay}, \citenamefont
  {Mani},\ and\ \citenamefont {Angom}}]{chattopadhyay-12b}%
  \BibitemOpen
  \bibfield  {author} {\bibinfo {author} {\bibfnamefont {S.}~\bibnamefont
  {Chattopadhyay}}, \bibinfo {author} {\bibfnamefont {B.~K.}\ \bibnamefont
  {Mani}}, \ and\ \bibinfo {author} {\bibfnamefont {D.}~\bibnamefont {Angom}},\
  }\bibfield  {title} {\enquote {\bibinfo {title} {Perturbed coupled-cluster
  theory to calculate dipole polarizabilities of closed-shell systems:
  Application to ar, kr, xe, and rn},}\ }\href {\doibase
  10.1103/PhysRevA.86.062508} {\bibfield  {journal} {\bibinfo  {journal} {Phys.
  Rev. A}\ }\textbf {\bibinfo {volume} {86}},\ \bibinfo {pages} {062508}
  (\bibinfo {year} {2012}{\natexlab{a}})}\BibitemShut {NoStop}%
\bibitem [{\citenamefont {Chattopadhyay}\ \emph
  {et~al.}(2013{\natexlab{a}})\citenamefont {Chattopadhyay}, \citenamefont
  {Mani},\ and\ \citenamefont {Angom}}]{chattopadhyay-13b}%
  \BibitemOpen
  \bibfield  {author} {\bibinfo {author} {\bibfnamefont {S.}~\bibnamefont
  {Chattopadhyay}}, \bibinfo {author} {\bibfnamefont {B.~K.}\ \bibnamefont
  {Mani}}, \ and\ \bibinfo {author} {\bibfnamefont {D.}~\bibnamefont {Angom}},\
  }\bibfield  {title} {\enquote {\bibinfo {title} {Electric dipole
  polarizabilities of doubly ionized alkaline-earth-metal ions from perturbed
  relativistic coupled-cluster theory},}\ }\href {\doibase
  10.1103/PhysRevA.87.062504} {\bibfield  {journal} {\bibinfo  {journal} {Phys.
  Rev. A}\ }\textbf {\bibinfo {volume} {87}},\ \bibinfo {pages} {062504}
  (\bibinfo {year} {2013}{\natexlab{a}})}\BibitemShut {NoStop}%
\bibitem [{\citenamefont {Chattopadhyay}\ \emph {et~al.}(2014)\citenamefont
  {Chattopadhyay}, \citenamefont {Mani},\ and\ \citenamefont
  {Angom}}]{chattopadhyay-14}%
  \BibitemOpen
  \bibfield  {author} {\bibinfo {author} {\bibfnamefont {S.}~\bibnamefont
  {Chattopadhyay}}, \bibinfo {author} {\bibfnamefont {B.~K.}\ \bibnamefont
  {Mani}}, \ and\ \bibinfo {author} {\bibfnamefont {D.}~\bibnamefont {Angom}},\
  }\bibfield  {title} {\enquote {\bibinfo {title} {Electric dipole
  polarizability of alkaline-earth-metal atoms from perturbed relativistic
  coupled-cluster theory with triples},}\ }\href@noop {} {\bibfield  {journal}
  {\bibinfo  {journal} {Phys. Rev. A}\ }\textbf {\bibinfo {volume} {89}},\
  \bibinfo {pages} {022506} (\bibinfo {year} {2014})}\BibitemShut {NoStop}%
\bibitem [{\citenamefont {Chattopadhyay}\ \emph {et~al.}(2015)\citenamefont
  {Chattopadhyay}, \citenamefont {Mani},\ and\ \citenamefont
  {Angom}}]{chattopadhyay-15}%
  \BibitemOpen
  \bibfield  {author} {\bibinfo {author} {\bibfnamefont {S.}~\bibnamefont
  {Chattopadhyay}}, \bibinfo {author} {\bibfnamefont {B.~K.}\ \bibnamefont
  {Mani}}, \ and\ \bibinfo {author} {\bibfnamefont {D.}~\bibnamefont {Angom}},\
  }\bibfield  {title} {\enquote {\bibinfo {title} {Triple excitations in
  perturbed relativistic coupled-cluster theory and electric dipole
  polarizability of groupiib elements},}\ }\href@noop {} {\bibfield  {journal}
  {\bibinfo  {journal} {Phys. Rev. A}\ }\textbf {\bibinfo {volume} {91}},\
  \bibinfo {pages} {052504} (\bibinfo {year} {2015})}\BibitemShut {NoStop}%
\bibitem [{\citenamefont {Kumar}\ \emph {et~al.}(2020)\citenamefont {Kumar},
  \citenamefont {Chattopadhyay}, \citenamefont {Mani},\ and\ \citenamefont
  {Angom}}]{ravi-20}%
  \BibitemOpen
  \bibfield  {author} {\bibinfo {author} {\bibfnamefont {Ravi}\ \bibnamefont
  {Kumar}}, \bibinfo {author} {\bibfnamefont {S.}~\bibnamefont
  {Chattopadhyay}}, \bibinfo {author} {\bibfnamefont {B.~K.}\ \bibnamefont
  {Mani}}, \ and\ \bibinfo {author} {\bibfnamefont {D.}~\bibnamefont {Angom}},\
  }\bibfield  {title} {\enquote {\bibinfo {title} {Electric dipole
  polarizability of group-13 ions using perturbed relativistic coupled-cluster
  theory: Importance of nonlinear terms},}\ }\href {\doibase
  10.1103/PhysRevA.101.012503} {\bibfield  {journal} {\bibinfo  {journal}
  {Phys. Rev. A}\ }\textbf {\bibinfo {volume} {101}},\ \bibinfo {pages}
  {012503} (\bibinfo {year} {2020})}\BibitemShut {NoStop}%
\bibitem [{\citenamefont {Safronova}\ \emph {et~al.}(1999)\citenamefont
  {Safronova}, \citenamefont {Johnson},\ and\ \citenamefont
  {Derevianko}}]{safronova-99}%
  \BibitemOpen
  \bibfield  {author} {\bibinfo {author} {\bibfnamefont {M.~S.}\ \bibnamefont
  {Safronova}}, \bibinfo {author} {\bibfnamefont {W.~R.}\ \bibnamefont
  {Johnson}}, \ and\ \bibinfo {author} {\bibfnamefont {A.}~\bibnamefont
  {Derevianko}},\ }\bibfield  {title} {\enquote {\bibinfo {title} {Relativistic
  many-body calculations of energy levels, hyperfine constants, electric-dipole
  matrix elements, and static polarizabilities for alkali-metal atoms},}\
  }\href {\doibase 10.1103/PhysRevA.60.4476} {\bibfield  {journal} {\bibinfo
  {journal} {Phys. Rev. A}\ }\textbf {\bibinfo {volume} {60}},\ \bibinfo
  {pages} {4476--4487} (\bibinfo {year} {1999})}\BibitemShut {NoStop}%
\bibitem [{\citenamefont {Derevianko}\ \emph {et~al.}(1999)\citenamefont
  {Derevianko}, \citenamefont {Johnson}, \citenamefont {Safronova},\ and\
  \citenamefont {Babb}}]{derevianko-99}%
  \BibitemOpen
  \bibfield  {author} {\bibinfo {author} {\bibfnamefont {A.}~\bibnamefont
  {Derevianko}}, \bibinfo {author} {\bibfnamefont {W.~R.}\ \bibnamefont
  {Johnson}}, \bibinfo {author} {\bibfnamefont {M.~S.}\ \bibnamefont
  {Safronova}}, \ and\ \bibinfo {author} {\bibfnamefont {J.~F.}\ \bibnamefont
  {Babb}},\ }\bibfield  {title} {\enquote {\bibinfo {title} {High-precision
  calculations of dispersion coefficients, static dipole polarizabilities, and
  atom-wall interaction constants for alkali-metal atoms},}\ }\href {\doibase
  10.1103/PhysRevLett.82.3589} {\bibfield  {journal} {\bibinfo  {journal}
  {Phys. Rev. Lett.}\ }\textbf {\bibinfo {volume} {82}},\ \bibinfo {pages}
  {3589--3592} (\bibinfo {year} {1999})}\BibitemShut {NoStop}%
\bibitem [{\citenamefont {Sucher}(1980)}]{sucher-80}%
  \BibitemOpen
  \bibfield  {author} {\bibinfo {author} {\bibfnamefont {J.}~\bibnamefont
  {Sucher}},\ }\bibfield  {title} {\enquote {\bibinfo {title} {Foundations of
  the relativistic theory of many-electron atoms},}\ }\href {\doibase
  10.1103/PhysRevA.22.348} {\bibfield  {journal} {\bibinfo  {journal} {Phys.
  Rev. A}\ }\textbf {\bibinfo {volume} {22}},\ \bibinfo {pages} {348--362}
  (\bibinfo {year} {1980})}\BibitemShut {NoStop}%
\bibitem [{\citenamefont {Mohanty}\ \emph {et~al.}(1991)\citenamefont
  {Mohanty}, \citenamefont {Parpia},\ and\ \citenamefont
  {Clementi}}]{mohanty-91}%
  \BibitemOpen
  \bibfield  {author} {\bibinfo {author} {\bibfnamefont {A.~K.}\ \bibnamefont
  {Mohanty}}, \bibinfo {author} {\bibfnamefont {F.~A.}\ \bibnamefont {Parpia}},
  \ and\ \bibinfo {author} {\bibfnamefont {E.}~\bibnamefont {Clementi}},\
  }\bibfield  {title} {\enquote {\bibinfo {title} {Kinetically balanced
  geometric gaussian basis set calculations for relativistic many-electron
  atoms},}\ }in\ \href@noop {} {\emph {\bibinfo {booktitle} {Modern Techniques
  in Computational Chemistry: MOTECC-91}}},\ \bibinfo {editor} {edited by\
  \bibinfo {editor} {\bibfnamefont {E.}~\bibnamefont {Clementi}}}\ (\bibinfo
  {publisher} {ESCOM},\ \bibinfo {year} {1991})\BibitemShut {NoStop}%
\bibitem [{\citenamefont {Stanton}\ and\ \citenamefont
  {Havriliak}(1984)}]{stanton-84}%
  \BibitemOpen
  \bibfield  {author} {\bibinfo {author} {\bibfnamefont {Richard~E.}\
  \bibnamefont {Stanton}}\ and\ \bibinfo {author} {\bibfnamefont {Stephen}\
  \bibnamefont {Havriliak}},\ }\bibfield  {title} {\enquote {\bibinfo {title}
  {Kinetic balance: A partial solution to the problem of variational safety in
  dirac calculations},}\ }\href {\doibase 10.1063/1.447865} {\bibfield
  {journal} {\bibinfo  {journal} {J. Chem. Phys.}\ }\textbf {\bibinfo {volume}
  {81}},\ \bibinfo {pages} {1910--1918} (\bibinfo {year} {1984})}\BibitemShut
  {NoStop}%
\bibitem [{\citenamefont {Grant}(2010)}]{grant-10}%
  \BibitemOpen
  \bibfield  {author} {\bibinfo {author} {\bibfnamefont {I.~P.}\ \bibnamefont
  {Grant}},\ }\href@noop {} {\emph {\bibinfo {title} {Relativistic Quantum
  Theory of Atoms and Molecules: Theory and Computation}}}\ (\bibinfo
  {publisher} {Springer},\ \bibinfo {address} {New York},\ \bibinfo {year}
  {2010})\BibitemShut {NoStop}%
\bibitem [{\citenamefont {Grant}(2006)}]{grant-06}%
  \BibitemOpen
  \bibfield  {author} {\bibinfo {author} {\bibfnamefont {Ian}\ \bibnamefont
  {Grant}},\ }\bibfield  {title} {\enquote {\bibinfo {title} {Relativistic
  atomic structure},}\ }in\ \href {\doibase 10.1007/978-0-387-26308-3_22}
  {\emph {\bibinfo {booktitle} {Springer Handbook of Atomic, Molecular, and
  Optical Physics}}},\ \bibinfo {editor} {edited by\ \bibinfo {editor}
  {\bibfnamefont {Gordon}\ \bibnamefont {Drake}}}\ (\bibinfo  {publisher}
  {Springer},\ \bibinfo {address} {New York},\ \bibinfo {year} {2006})\ pp.\
  \bibinfo {pages} {325--357}\BibitemShut {NoStop}%
\bibitem [{\citenamefont {Chattopadhyay}\ \emph
  {et~al.}(2012{\natexlab{b}})\citenamefont {Chattopadhyay}, \citenamefont
  {Mani},\ and\ \citenamefont {Angom}}]{chattopadhyay-12a}%
  \BibitemOpen
  \bibfield  {author} {\bibinfo {author} {\bibfnamefont {S.}~\bibnamefont
  {Chattopadhyay}}, \bibinfo {author} {\bibfnamefont {B.~K.}\ \bibnamefont
  {Mani}}, \ and\ \bibinfo {author} {\bibfnamefont {D.}~\bibnamefont {Angom}},\
  }\bibfield  {title} {\enquote {\bibinfo {title} {Electric dipole
  polarizability from perturbed relativistic coupled-cluster theory:
  Application to neon},}\ }\href {\doibase 10.1103/PhysRevA.86.022522}
  {\bibfield  {journal} {\bibinfo  {journal} {Phys. Rev. A}\ }\textbf {\bibinfo
  {volume} {86}},\ \bibinfo {pages} {022522} (\bibinfo {year}
  {2012}{\natexlab{b}})}\BibitemShut {NoStop}%
\bibitem [{\citenamefont {Chattopadhyay}\ \emph
  {et~al.}(2013{\natexlab{b}})\citenamefont {Chattopadhyay}, \citenamefont
  {Mani},\ and\ \citenamefont {Angom}}]{chattopadhyay-13a}%
  \BibitemOpen
  \bibfield  {author} {\bibinfo {author} {\bibfnamefont {S.}~\bibnamefont
  {Chattopadhyay}}, \bibinfo {author} {\bibfnamefont {B.~K.}\ \bibnamefont
  {Mani}}, \ and\ \bibinfo {author} {\bibfnamefont {D.}~\bibnamefont {Angom}},\
  }\bibfield  {title} {\enquote {\bibinfo {title} {Electric dipole
  polarizabilities of alkali-metal ions from perturbed relativistic
  coupled-cluster theory},}\ }\href {\doibase 10.1103/PhysRevA.87.042520}
  {\bibfield  {journal} {\bibinfo  {journal} {Phys. Rev. A}\ }\textbf {\bibinfo
  {volume} {87}},\ \bibinfo {pages} {042520} (\bibinfo {year}
  {2013}{\natexlab{b}})}\BibitemShut {NoStop}%
\bibitem [{\citenamefont {Purvis}\ and\ \citenamefont
  {Bartlett}(1982)}]{purvis-82}%
  \BibitemOpen
  \bibfield  {author} {\bibinfo {author} {\bibfnamefont {George~D.}\
  \bibnamefont {Purvis}}\ and\ \bibinfo {author} {\bibfnamefont {Rodney~J.}\
  \bibnamefont {Bartlett}},\ }\bibfield  {title} {\enquote {\bibinfo {title} {A
  full coupled‐cluster singles and doubles model: The inclusion of
  disconnected triples},}\ }\href {\doibase http://dx.doi.org/10.1063/1.443164}
  {\bibfield  {journal} {\bibinfo  {journal} {J. Chem. Phys.}\ }\textbf
  {\bibinfo {volume} {76}},\ \bibinfo {pages} {1910--1918} (\bibinfo {year}
  {1982})}\BibitemShut {NoStop}%
\bibitem [{\citenamefont {Mani}\ and\ \citenamefont {Angom}(2010)}]{mani-10}%
  \BibitemOpen
  \bibfield  {author} {\bibinfo {author} {\bibfnamefont {B.~K.}\ \bibnamefont
  {Mani}}\ and\ \bibinfo {author} {\bibfnamefont {D.}~\bibnamefont {Angom}},\
  }\bibfield  {title} {\enquote {\bibinfo {title} {Atomic properties calculated
  by relativistic coupled-cluster theory without truncation: Hyperfine
  constants of ${\mathrm{mg}}^{+}$, ${\mathrm{ca}}^{+}$, ${\mathrm{sr}}^{+}$,
  and ${\mathrm{ba}}^{+}$},}\ }\href {\doibase 10.1103/PhysRevA.81.042514}
  {\bibfield  {journal} {\bibinfo  {journal} {Phys. Rev. A}\ }\textbf {\bibinfo
  {volume} {81}},\ \bibinfo {pages} {042514} (\bibinfo {year}
  {2010})}\BibitemShut {NoStop}%
\bibitem [{\citenamefont {J{\"{o}}nsson}\ \emph {et~al.}(2013)\citenamefont
  {J{\"{o}}nsson}, \citenamefont {Gaigalas}, \citenamefont {Biero{\'{n}}},
  \citenamefont {Froese~Fischer},\ and\ \citenamefont {Grant}}]{jonsson-13}%
  \BibitemOpen
  \bibfield  {author} {\bibinfo {author} {\bibfnamefont {P.}~\bibnamefont
  {J{\"{o}}nsson}}, \bibinfo {author} {\bibfnamefont {G.}~\bibnamefont
  {Gaigalas}}, \bibinfo {author} {\bibfnamefont {J.}~\bibnamefont
  {Biero{\'{n}}}}, \bibinfo {author} {\bibfnamefont {C.}~\bibnamefont
  {Froese~Fischer}}, \ and\ \bibinfo {author} {\bibfnamefont {I.~P.}\
  \bibnamefont {Grant}},\ }\bibfield  {title} {\enquote {\bibinfo {title} {New
  version: Grasp2k relativistic atomic structure package},}\ }\href {\doibase
  http://dx.doi.org/10.1016/j.cpc.2013.02.016} {\bibfield  {journal} {\bibinfo
  {journal} {Comp. Phys. Comm.}\ }\textbf {\bibinfo {volume} {184}},\ \bibinfo
  {pages} {2197 -- 2203} (\bibinfo {year} {2013})}\BibitemShut {NoStop}%
\bibitem [{\citenamefont {Zatsarinny}\ and\ \citenamefont {{Froese
  Fischer}}(2016)}]{zatsarinny-16}%
  \BibitemOpen
  \bibfield  {author} {\bibinfo {author} {\bibfnamefont {Oleg}\ \bibnamefont
  {Zatsarinny}}\ and\ \bibinfo {author} {\bibfnamefont {Charlotte}\
  \bibnamefont {{Froese Fischer}}},\ }\bibfield  {title} {\enquote {\bibinfo
  {title} {{DBSR-HF: A B-spline Dirac–Hartree–Fock program}},}\ }\href
  {\doibase https://doi.org/10.1016/j.cpc.2015.12.023} {\bibfield  {journal}
  {\bibinfo  {journal} {Computer Physics Communications}\ }\textbf {\bibinfo
  {volume} {202}},\ \bibinfo {pages} {287 -- 303} (\bibinfo {year}
  {2016})}\BibitemShut {NoStop}%
\bibitem [{\citenamefont {Kozioł}\ and\ \citenamefont
  {Aucar}(2018)}]{karol-18b}%
  \BibitemOpen
  \bibfield  {author} {\bibinfo {author} {\bibfnamefont {Karol}\ \bibnamefont
  {Kozioł}}\ and\ \bibinfo {author} {\bibfnamefont {Gustavo~A.}\ \bibnamefont
  {Aucar}},\ }\bibfield  {title} {\enquote {\bibinfo {title} {Qed effects on
  individual atomic orbital energies},}\ }\href {\doibase 10.1063/1.5026193}
  {\bibfield  {journal} {\bibinfo  {journal} {The Journal of Chemical Physics}\
  }\textbf {\bibinfo {volume} {148}},\ \bibinfo {pages} {134101} (\bibinfo
  {year} {2018})},\ \Eprint
  {http://arxiv.org/abs/https://doi.org/10.1063/1.5026193}
  {https://doi.org/10.1063/1.5026193} \BibitemShut {NoStop}%
\bibitem [{\citenamefont {Mani}\ \emph {et~al.}(2017)\citenamefont {Mani},
  \citenamefont {Chattopadhyay},\ and\ \citenamefont {Angom}}]{mani-17}%
  \BibitemOpen
  \bibfield  {author} {\bibinfo {author} {\bibfnamefont {B.K.}\ \bibnamefont
  {Mani}}, \bibinfo {author} {\bibfnamefont {S.}~\bibnamefont {Chattopadhyay}},
  \ and\ \bibinfo {author} {\bibfnamefont {D.}~\bibnamefont {Angom}},\
  }\bibfield  {title} {\enquote {\bibinfo {title} {Rccpac: A parallel
  relativistic coupled-cluster program for closed-shell and one-valence atoms
  and ions in fortran},}\ }\href {\doibase
  https://doi.org/10.1016/j.cpc.2016.11.008} {\bibfield  {journal} {\bibinfo
  {journal} {Computer Physics Communications}\ }\textbf {\bibinfo {volume}
  {213}},\ \bibinfo {pages} {136 -- 154} (\bibinfo {year} {2017})}\BibitemShut
  {NoStop}%
\bibitem [{\citenamefont {Flores}\ and\ \citenamefont
  {Redondo}(1993)}]{flores-93a}%
  \BibitemOpen
  \bibfield  {author} {\bibinfo {author} {\bibfnamefont {J~R}\ \bibnamefont
  {Flores}}\ and\ \bibinfo {author} {\bibfnamefont {P}~\bibnamefont
  {Redondo}},\ }\bibfield  {title} {\enquote {\bibinfo {title} {Computation of
  second-order correlation energies using a finite element method for atoms
  with d electrons},}\ }\href {\doibase 10.1088/0953-4075/26/15/012} {\bibfield
   {journal} {\bibinfo  {journal} {Journal of Physics B: Atomic, Molecular and
  Optical Physics}\ }\textbf {\bibinfo {volume} {26}},\ \bibinfo {pages}
  {2251--2261} (\bibinfo {year} {1993})}\BibitemShut {NoStop}%
\bibitem [{\citenamefont {Flores}\ and\ \citenamefont
  {Redondo}(1994)}]{flores-94}%
  \BibitemOpen
  \bibfield  {author} {\bibinfo {author} {\bibfnamefont {Jesús~R.}\
  \bibnamefont {Flores}}\ and\ \bibinfo {author} {\bibfnamefont
  {P.}~\bibnamefont {Redondo}},\ }\bibfield  {title} {\enquote {\bibinfo
  {title} {High-precision atomic computations from finite element techniques:
  Second-order correlation energies for be, ca, sr, cd, ba, yb, and hg},}\
  }\href {\doibase 10.1002/jcc.540150710} {\bibfield  {journal} {\bibinfo
  {journal} {Journal of Computational Chemistry}\ }\textbf {\bibinfo {volume}
  {15}},\ \bibinfo {pages} {782--790} (\bibinfo {year} {1994})},\ \Eprint
  {http://arxiv.org/abs/https://onlinelibrary.wiley.com/doi/pdf/10.1002/jcc.540150710}
  {https://onlinelibrary.wiley.com/doi/pdf/10.1002/jcc.540150710} \BibitemShut
  {NoStop}%
\bibitem [{\citenamefont {Flores}(1993)}]{flores-93b}%
  \BibitemOpen
  \bibfield  {author} {\bibinfo {author} {\bibfnamefont {Jesús~R.}\
  \bibnamefont {Flores}},\ }\bibfield  {title} {\enquote {\bibinfo {title}
  {High precision atomic computations from finite element techniques:
  Second‐order correlation energies of rare gas atoms},}\ }\href {\doibase
  10.1063/1.464908} {\bibfield  {journal} {\bibinfo  {journal} {The Journal of
  Chemical Physics}\ }\textbf {\bibinfo {volume} {98}},\ \bibinfo {pages}
  {5642--5647} (\bibinfo {year} {1993})},\ \Eprint
  {http://arxiv.org/abs/https://doi.org/10.1063/1.464908}
  {https://doi.org/10.1063/1.464908} \BibitemShut {NoStop}%
\bibitem [{\citenamefont {Ishikawa}\ and\ \citenamefont
  {Koc}(1994)}]{ishikawa-94}%
  \BibitemOpen
  \bibfield  {author} {\bibinfo {author} {\bibfnamefont {Yasuyuki}\
  \bibnamefont {Ishikawa}}\ and\ \bibinfo {author} {\bibfnamefont {Konrad}\
  \bibnamefont {Koc}},\ }\bibfield  {title} {\enquote {\bibinfo {title}
  {Relativistic many-body perturbation theory based on the no-pair
  {Dirac-Coulomb-Breit Hamiltonian}: {R}elativistic correlation energies for
  the noble-gas sequence through {Rn (Z=86)}, the group-iib atoms through {Hg},
  and the ions of {Ne} isoelectronic sequence},}\ }\href {\doibase
  10.1103/PhysRevA.50.4733} {\bibfield  {journal} {\bibinfo  {journal} {Phys.
  Rev. A}\ }\textbf {\bibinfo {volume} {50}},\ \bibinfo {pages} {4733--4742}
  (\bibinfo {year} {1994})}\BibitemShut {NoStop}%
\bibitem [{\citenamefont {Mani}\ \emph {et~al.}(2009)\citenamefont {Mani},
  \citenamefont {Latha},\ and\ \citenamefont {Angom}}]{mani-09}%
  \BibitemOpen
  \bibfield  {author} {\bibinfo {author} {\bibfnamefont {B.~K.}\ \bibnamefont
  {Mani}}, \bibinfo {author} {\bibfnamefont {K.~V.~P.}\ \bibnamefont {Latha}},
  \ and\ \bibinfo {author} {\bibfnamefont {D.}~\bibnamefont {Angom}},\
  }\bibfield  {title} {\enquote {\bibinfo {title} {Relativistic coupled-cluster
  calculations of $^{20}\text{N}\text{e}$, $^{40}\text{A}\text{r}$,
  $^{84}\text{K}\text{r}$, and $^{129}\text{X}\text{e}$: Correlation energies
  and dipole polarizabilities},}\ }\href {\doibase 10.1103/PhysRevA.80.062505}
  {\bibfield  {journal} {\bibinfo  {journal} {Phys. Rev. A}\ }\textbf {\bibinfo
  {volume} {80}},\ \bibinfo {pages} {062505} (\bibinfo {year}
  {2009})}\BibitemShut {NoStop}%
\bibitem [{\citenamefont {McCarthy}\ and\ \citenamefont
  {Thakkar}(2011)}]{mccarthy-11}%
  \BibitemOpen
  \bibfield  {author} {\bibinfo {author} {\bibfnamefont {Shane~P.}\
  \bibnamefont {McCarthy}}\ and\ \bibinfo {author} {\bibfnamefont {Ajit~J.}\
  \bibnamefont {Thakkar}},\ }\bibfield  {title} {\enquote {\bibinfo {title}
  {Accurate all-electron correlation energies for the closed-shell atoms from
  {A}r to {R}n and their relationship to the corresponding {MP2} correlation
  energies},}\ }\href {\doibase 10.1063/1.3547262} {\bibfield  {journal}
  {\bibinfo  {journal} {The Journal of Chemical Physics}\ }\textbf {\bibinfo
  {volume} {134}},\ \bibinfo {pages} {044102} (\bibinfo {year} {2011})},\
  \Eprint {http://arxiv.org/abs/https://doi.org/10.1063/1.3547262}
  {https://doi.org/10.1063/1.3547262} \BibitemShut {NoStop}%
\end{thebibliography}%

\end{document}